\documentclass[useAMS,usenatbib]{mn2e} 

\usepackage{times}
\usepackage{graphicx}
\usepackage{amssymb}
\newcommand{\HI}{H\,{\sevensize I}}

\newcommand{\NHI}{$N\rm _{H\,{\sevensize I}}$}

\newcommand{\CIIs}{[C\,{\sevensize II}*]}
\newcommand{\CII}{C\,{\sevensize II}}
\newcommand{\lya}{Ly$\alpha$}

\newcommand{\sfr}{M$_{\odot}$ yr$^{-1}$}
\newcommand{\Ssfr}{M$_{\odot}$ yr$^{-1}$ kpc$^{-2}$}
\newcommand{\cmm}{cm$^{-2}$}

\title[Imaging DLAs at $z>2$ (III)]{Directly imaging damped Ly-$\alpha$ 
galaxies at $z>2$. III: The star formation rates of neutral gas reservoirs at $z\sim2.7$}

\author[Fumagalli et al.]{Michele Fumagalli$^{1,2}$\thanks{E-mail: michele.fumagalli@durham.ac.uk}, 
 John M. O'Meara$^{3}$, J. Xavier Prochaska$^{4,5}$,  Marc Rafelski$^6$, \and Nissim Kanekar$^{7}$ \\
$^{1}$Institute for Computational Cosmology, Department of Physics, Durham University, 
South Road, Durham, DH1 3LE, UK \\
$^{2}$Carnegie Observatories, 813 Santa Barbara Street, Pasadena, CA 91101, USA \\
$^{3}$Department of Chemistry and Physics, Saint Michael's College, One Winooski Park, 
     Colchester, VT 05439, USA \\
$^{4}$Department of Astronomy and Astrophysics, University of California, 1156 High Street, 
     Santa Cruz, CA 95064 USA \\
$^{5}$University of California Observatories, Lick Observatory 1156 High Street, Santa Cruz, CA 95064 USA \\
$^{6}$NASA Postdoctoral Program Fellow, Goddard Space Flight Center, Code 665, Greenbelt, MD 20771, USA \\
$^{7}$National Centre for Radio Astrophysics, TIFR, Post Bag 3, Ganeshkhind, Pune 411 007, India}

\begin{document}

\date{Accepted xxxx. Received xxxx; in original form xxxx}

\pagerange{\pageref{firstpage}--\pageref{lastpage}} \pubyear{xxxx}

\maketitle

\label{firstpage}

\begin{abstract}
We present results from a survey designed to probe the star formation properties of 32
damped Ly$\alpha$ systems (DLAs) at $z\sim 2.7$. By using the ``double-DLA'' technique
that eliminates the glare of the bright background quasars, 
we directly measure the rest-frame FUV flux from DLAs and their neighbouring galaxies. 
At the position of the absorbing gas, 
we place stringent constraints on the unobscured star formation rates (SFRs) of 
DLAs to $2\sigma$ limits of $\dot\psi<0.09-0.27~$\sfr, 
corresponding to SFR surface densities $\Sigma_{\rm sfr}<10^{-2.6}-10^{-1.5}~$\Ssfr.
The implications of these limits for the star formation law, metal enrichment, 
and cooling rates of DLAs are examined. By studying the distribution of impact parameters 
as a function of SFRs for all the galaxies detected around these DLAs, 
we place new direct constraints on the bright end of the UV luminosity function of DLA hosts. 
We find that $\le 13\%$ of the hosts have $\dot\psi \ge 2$\sfr\ at impact 
parameters $b_{\rm dla} \le (\dot\psi/{\rm M_\odot~yr^{-1}})^{0.8}+6~\rm kpc$, differently from 
current samples of confirmed DLA galaxies.
Our observations also disfavor a scenario in which the majority of 
DLAs arise from bright LBGs at distances $20 \le b_{\rm dla} < 100~\rm kpc$.
These new findings corroborate a picture in which DLAs do not originate from  
highly star forming systems that are coincident with the absorbers, and instead 
suggest that DLAs are associated with faint, possibly isolated, star-forming galaxies. 
Potential shortcomings of this scenario and future strategies for further investigation are discussed. 
\end{abstract}


%--------------------------------------
\section{Introduction}

The continuous improvement of the instrumentation available at ground based observatories 
and on board of the {\it Hubble Space Telescope} (HST) has facilitated numerous deep galaxy 
surveys across wide areas of the sky to characterize the typical star formation rates (SFRs), 
dust content, stellar masses, and structural properties of galaxies across cosmic time, up to 
the epoch of reionization. However, the redshift-dependent sensitivity of these surveys 
offers only a view of the tip of the iceberg of the galaxy population in the distant 
Universe. An appealing alternative to explore the physical properties of galaxies across 
a wider range of luminosity and masses is the study of damped Lyman-$\alpha$ absorbers 
\citep[DLAs;][]{wol86,wol05} that are detected along the line of sight to 
bright background sources such as quasars or the afterglows of $\gamma$-ray bursts (GRBs). 
Being selected solely according to their neutral hydrogen column density 
(\NHI$\ge 2\times 10^{20}$ \cmm), DLAs arise in dense and neutral gas and, as suggested 
also by models and hydrodynamic simulations, they appear to trace the hydrogen content of galaxies 
over a wide range of halo masses \citep[e.g.][]{nag07,pon08,tes09,hon10,cen12,bir14}.

The study of hydrogen absorption lines in large spectroscopic samples of DLAs 
\citep[e.g.][]{pro05,pro09,not09,not12,zaf13} and of the associated metal lines in 
high-resolution spectra \citep[e.g][]{led06,pro07,raf12,jor13,mol13,nee13,raf14} have constrained 
the kinematics and metallicities of the gas in these absorbers at different redshifts. 
DLAs at redshifts $z=2-3$ cover roughly one third of the sky and evolve in time such 
that the shape of the column density distribution function is preserved 
(\citealt{zwa05,pro09}; but see \citealt{bra12}). 
They are typically sub-solar in composition, but more enriched ($\sim 5\%$ of the solar 
metallicity) than the intergalactic medium \citep[IGM;][]{pet97,pro07,fum11,raf12,jor13}. 
Finally, DLAs exhibit abundance patterns that can be described with bursty star formation 
histories that are common in irregular and dwarf galaxies in stochastic regimes \citep{des07,das14,dom14}, 
and, as suggested by recent studies, that follow the chemical composition observed in Galactic 
halo stars \citep{raf12} or local dwarfs \citep{coo14}.

However, an outstanding challenge that has limited our understanding of the nature
of DLAs, and their link to the galaxy population at high redshift, is the identification of the
host galaxies of the absorbers.  
In fact, the bright background light that enables absorption spectroscopy 
becomes a limiting source of contamination during imaging observations, hampering the 
identification of galaxies at close impact parameters to the quasars \citep[e.g.][]{mol98}. 
For this reason, despite over twenty years of investigation \citep[see Appendix B of][]{fum10b}, 
the emitting counterparts  of only a dozen DLAs have been detected to date at $z\gtrsim 2$ 
\citep[Table \ref{known}; see also][]{kro12}. 
Many of these detections have stemmed from recent spectroscopic and integral field unit (IFU) searches 
\citep{fyn10,fyn11,per12,not12b,fyn13,bou13,jor14,raf14}. 

Despite these recent successes, it is difficult to assess the extent to
which the identified DLA counterparts represent the general population of DLA host 
galaxies; this is due to both the selection criteria used in some searches (e.g. 
targeting high metallicity DLAs) and the absence of a reliable census of non-detections.
Furthermore, by design, 
the bright background source used for absorption spectroscopy is perfectly aligned to the absorbing 
material, making a direct study of the {\it in-situ} SFRs of DLAs detected against quasars
impossible even in surveys that target representative samples of DLAs at the high resolution offered by 
HST \citep[e.g.][]{war01}. In turn, this prevents a systematic analysis of the star formation law in neutral 
gas clouds using far ultraviolet (FUV) tracers in the distant Universe. And, in fact, previous work on the 
subject has been limited to \lya\ emission \citep{rah10,not14,cai14} or
indirect methods only, such as statistically connecting DLAs to low 
surface brightness galaxies \citep{wol06} and to the outskirts of compact Lyman break galaxies
\citep[LBGs;][]{raf11}.

Pressing questions on what the typical DLA counterparts are or at what rate 
these \HI\ clouds form stars remain open. Are DLA galaxies massive rotating 
disks as suggested by the absorption line profiles \citep[e.g.][]{pro97} or the 
small building blocks predicted by the theories of the hierarchical assembly of 
structures \citep[e.g.][]{hae98,rau08}? Or are DLAs gas clouds in the halo of star-forming 
galaxies \citep[e.g.][]{bou13}, possibly hosting vigorous starburst-driven outflows 
\citep[e.g.][]{not12b}? And what is the source of heating that balances the 
cooling rates estimated from the \CIIs\ absorption lines \citep{wol03,wol03a}? 

Recently, two promising methods to address some of the above questions have been explored.  
The first approach focuses on intervening DLAs detected along the line of sight to GRBs,
taking advantage of the fact that once the bright afterglow has faded, deep imaging 
follow-up can be conducted to search for DLA counterparts in emission at all impact 
parameters. The downside of this approach is that the transient nature of GRBs limits 
the size of the samples for which good-quality high-resolution spectra are available to study 
the DLA properties in absorption \citep[e.g.][]{cuc14}. At present, only a handful of GRB sightlines
with intervening DLAs have been searched for the optical counterparts of these absorbers \citep{sch12}.

The second approach is to exploit the fortuitous alignment of two optically-thick absorbers 
along individual quasars \citep{ste92,ome06,chr09}. The basic idea of this technique, which we dub 
the ``double-DLA'' or the ``Lyman limit'' technique, is to image quasar fields that host DLAs at lower redshift 
and optically-thick absorbers (DLAs or LLSs, that is Lyman limit systems) at higher redshift. 
As discussed in \citet{fum10b} and \citet{fum14}, this second higher-redshift 
system acts as a natural blocking filter that absorbs all the quasar light blueward of its 
Lyman limit, potentially allowing the detection of DLA counterparts at all impact 
parameters at these shorter wavelengths. 
Nevertheless, given the high intrinsic fluxes of the background quasars, positive 
detections should be carefully vetted to exclude residual contamination from the background 
sources \citep[see][Section \ref{ins:direct}]{cai14}.
The main limitation of this approach is that 
it requires a specific alignment of two absorbers both in space and redshift. However, 
thanks to the large volume probed by modern spectroscopic surveys \citep[e.g.][]{not12}, 
this experiment can be performed for several quasars with intervening DLAs at $z\gtrsim 2$.

\begin{table*}
\caption{Measured {\it in-situ} SFRs in different apertures centered at the quasar position for individual fields and for the composite images.}\label{tabinsfr}
\centering
\begin{tabular}{l c c c r r r r}
\hline
Field & $z_{\rm dla}$& $D_{\rm lum}$ & $\phi_{\rm igm}$ & $f_{\rm lim}(<0.25")$ & $\dot\psi(<0.25")$ &  $f_{\rm lim}(<1.5")$ & $\dot\psi(<1.5")$\\
  &  & (Mpc) & (mag) & ($\rm 10^{-31} erg~s^{-1} cm^{-2} Hz^{-1}$) & $\rm (M_\odot yr^{-1})$ & ($\rm 10^{-31} erg~s^{-1} cm^{-2} Hz^{-1}$) & $\rm (M_\odot yr^{-1})$\\
\hline
1:G1 &  2.9181 &    25152  &  1.43&-&-&$<$   6.85 & $<$   1.05\\
2:G2 &  2.6878 &    22760  &  0.62&-&-&$<$   4.53 & $<$   0.60\\
3:G3 &  2.3887 &    19706  &  0.18&-&-&   2.6$\pm$   1.0 &	0.3$\pm$   0.1\\
4:G4 &  2.6878 &    22760  &  0.62&-&-&$<$   4.92 & $<$   0.65\\
5:G5 &  2.5713 &    21563  &  0.20&-&-&  11.7$\pm$   2.6 &	1.4$\pm$   0.3\\
6:G6 &  3.7861 &    34422  &  1.80&-&-&$<$  16.43 & $<$   3.85\\
7:G7 &  2.7544 &    23449  &  0.80&-&-&$<$   4.70 & $<$   0.65\\
8:G9 &  2.7584 &    23490  &  0.81&-&-&$<$   3.49 & $<$   0.49\\
9:G10 &  2.4592 &   20420  &  0.23&-&-&$<$   1.84 & $<$   0.21\\
10:G11$^{*}$ &  3.5297 &  31646  &  0.78&-&-&   7.2$\pm$   2.3 &	1.5$\pm$   0.5\\
11:G12 &  2.6606 &  22480  &  0.56&-&-&$<$   3.93 & $<$   0.51\\
12:G13 &  2.5978 &  21835  &  0.43&-&-&$<$   3.29 & $<$   0.41\\
13:H1 &  2.7803 &   23717  &  0.23 & $<$   3.81 & $<$	0.54&-&-\\
14:H2 &  1.9127 &   14990  &  0.78 & $<$   3.25 & $<$	0.24&-&-\\
15:H3 &  3.2530 &   28685  &  1.08 & $<$   2.68 & $<$	0.49&-&-\\
16:H4 &  2.7067 &   22955  &  0.20 & $<$   3.76 & $<$	0.51&-&-\\
17:H5 &  3.0010 &   26021  &  0.32 & $<$   4.13 & $<$	0.66&-&-\\
18:H6 &  2.9586 &   25576  &  0.31 & $<$   3.04 & $<$	0.48&-&-\\
19:H7 &  3.3069 &   29259  &  0.40 & $<$   1.59 & $<$	0.30&-&-\\
20:H8 &  2.6320 &   22186  &  0.31 & $<$   2.28 & $<$	0.29&-&-\\
21:H9 &  1.8639 &   14518  &  0.58 & $<$   2.58 & $<$	0.18&-&-\\
22:H10 &  2.6826 &  22707  &  0.56 & $<$   3.36 & $<$	0.44&-&-\\
23:H11 &  3.2332 &  28474  &  1.03 & $<$   1.88 & $<$	0.34&-&-\\
24:H12 &  3.5629 &  32004  &  0.88 & $<$   2.99 & $<$	0.64&-&-\\
25:H13 &  2.6500 &  22371  &  0.39 & $<$   2.86 & $<$	0.37&-&-\\
26:H14 &  2.7333 &  23230  &  0.21 & $<$   2.79 & $<$	0.38&-&-\\
27:H15 &  2.9236 &  25210  &  0.30 & $<$   2.95 & $<$	0.45&-&-\\
28:H16 &  3.4035 &  30291  &  0.52 & $<$   1.81 & $<$	0.36&-&-\\
29:H17 &  2.6289 &  22154  &  0.30 & $<$   2.04 & $<$	0.26&-&-\\
30:H18 &  2.6079 &  21938  &  0.25 & $<$   2.34 & $<$	0.30&-&-\\
31:H19 &  2.8721 &  24672  &  0.41 & $<$   0.85 & $<$	0.13&-&-\\
32:H20 &  2.6722 &  22600  &  1.02 & $<$   3.21 & $<$	0.42&-&-\\
\hline
Stack & - & - & - & - & $<0.090$ & - & $<0.270$ \\
\hline
\end{tabular}
\flushleft{$^{*}$ For this sightline, we cannot exclude contamination from quasar light 
(see Section \ref{ins:direct} for details). The columns of the table are (1) name of the quasar field; (2) redshift of the DLA; 
(3) luminosity distance to the DLA; (4) correction for intervening IGM absorption; (5,7) limit on the FUV flux measured at the DLA position, within the aperture specified in parenthesis; (6,8) limit on the {\it in situ} SFR measured at the DLA position, within the aperture specified in parenthesis.}
\end{table*}

In previous papers of this series \citep{fum10b,fum14}, we discuss in detail
the survey rationale, and we demonstrate the power of this double-DLA technique in avoiding 
the contamination of the bright background quasars, which enables us to achieve the
same sensitivity limit at all impact parameters from the DLAs. In this previous work, 
we also describe the imaging and spectroscopic observations of the 32 quasar
fields with intervening DLAs at $z\sim 1.9-3.8$, which comprise our sample. 
Information on the data reduction, survey completeness limits, and preparation of the galaxy catalogues
can be found in \citet{fum14}, where we also present the \NHI\ and 
metallicity distribution of our sample, as measured in absorption with echellette
spectra. In our previous work, we also show that the targeted DLAs constitute an unbiased sample that 
represents the full range of absorption properties of the parent DLA population at these 
redshifts. 

In this third paper, we discuss the results of this survey. In Section 
\ref{insitu} we derive the first direct limits for the observed {\it in-situ} SFRs of DLAs in the UV, 
also discussing the implications for the metal content and the heating rates of DLA gas probed in absorption. 
In this section, we also compare our results to modern theories for star formation 
in atomic gas. Section \ref{dlagal} further extends our analysis beyond the inner few kpc where 
absorption arises, exploring the connection between DLAs and star-forming galaxies at any impact parameter.
We also compare our findings with previous observational studies.
In Section \ref{other}, we use our findings to inform the discussion of what DLAs are,
while a summary and conclusions follow in Section \ref{conclusion}. 

Throughout this work, consistent 
with our previous analysis, we express distances in proper units and magnitudes in the AB system, adopting 
the following cosmological parameters: $H_0=70.4~\rm km~s^{-1}~Mpc^{-1}$, $\Omega_{\rm m}=0.27$ and 
$\Omega_\Lambda=0.73$ \citep{kom11}. As in previous papers of this series, we refer to the absorption systems 
as ``DLA gas'' or simply DLAs, while we will explicitly refer to the host galaxies as ``DLA galaxy''.
Furthermore, we will often refer to two sub-samples: the first one, the ``HST sample'', includes 
20 DLAs which have been observed with WFC3 on board of HST; the second one, the ``ground-based sample'',
includes 12 DLAs which have been observed from the ground at Keck or at the Large Binocular Telescope. 
Details on the properties of these two sub-samples can be found in \citet{fum14}.
   
\section{The in-situ star formation of DLA\lowercase{s}}\label{insitu}

In this section, we take advantage of the unique possibility offered by the adopted 
double-DLA technique to explore, for the first time with direct measurements in the rest-frame UV, 
the star-formation activity associated with the absorbing DLA gas. 
We start by analyzing emission properties {\it in-situ},
that is at the position of the absorbing gas, inside individual DLAs (Section \ref{ins:direct}). 
We then generate composite images to extend our study to fainter limits (Section \ref{ins:stack}). 
Next, we connect these direct measurements to the absorption properties, discussing in turn 
the implications for the star formation law (Section \ref{ins:sflaw}),
the metal enrichment (Section \ref{ins:metal}), and the cooling rates
(Section \ref{sec:cool}) of DLAs. 

Before proceeding, we emphasize that our observations probe  FUV emission; thus, the measured SFRs  
represent only the fraction of the intrinsic star formation that originates in unobscured regions. 
In principle, we cannot rule out the existence of obscured star formation, but reddening measurements 
against quasars hosting DLAs imply the presence of only a modest amount of dust, with mean dust-to-gas 
ratios $A_V/N_{\rm HI} \sim 2-4 \times 10^{-23}~\rm mag~cm^{-2}$ \citep{vla08,kha12}. 
In the following, we will quote observed fluxes, non corrected for dust extinction. To help gauge the effects of 
dust extinction in our sample, we note that dust obscuration at the DLA position should lie in the range 
$A_V\sim 0.004-0.06$ mag in the DLA rest-frame. For an SMC-type extinction curve, this 
translates into a flux correction between $2-40\%$ at 1200\AA, depending on the gas column density. 
This calculation reassures us that, at least to first order, the {\it in-situ} SFR measurements of this section are 
representative of the bulk of the ongoing star formation in DLAs. However, as we will discuss in more detail below, 
our measurements average emission from regions that are more extended than the typical scales probed by 
absorption. Therefore, we cannot fully exclude the presence of clumps of obscured star formation
which may lie in close proximity to the quasar sightline.

\subsection{The star formation rate in DLA gas}

\subsubsection{SFRs in individual systems}\label{ins:direct}

We start by integrating the flux within two apertures of diameters 
$\theta = 0.25''$ and $\theta= 1.5''$ in each image. These apertures
represent physical sizes of $\sim 2$ and $\sim 12$ kpc at the median redshift of the DLA 
sample\footnote{For the adopted cosmology, $1''$ corresponds to 8.1 kpc at $z=2.7$.}.
These two apertures probe a population of DLAs that form stars {\it in situ}, either 
within compact or extended regions. We emphasize that in our analysis we 
consider the  background quasar position as a reference, although the location of a
sightline may not coincide with the geometric center of the physical structure that gives 
rise to the DLA. The sky level is computed locally in annuli centered at 
the DLA positions, extending between $4''-14''$ in radius. For the ground-based imaging, we compute SFRs only 
within the large aperture, which is more extended than the size of a point source. Conversely, 
we refrain from computing limits in the larger aperture for the HST sample, because, at the depth 
of the HST images, these limits are not informative compared to those obtained from the 
ground-based sample. These integrated fluxes, $f_{\rm int}$, are then compared to the 
flux upper limits, $f_{\rm ul}$, at $2\sigma$ confidence level (C.L.), which we compute by integrating the 
error models discussed in \citet{fum14} inside the same apertures. The final flux limits are set 
to be $f_{\rm lim} \equiv \max (f_{\rm int},f_{\rm ul})$. Lacking spectroscopic information, this 
definition allows us to express the most conservative upper limit on the {\it in-situ} 
SFRs of DLAs by treating positive detections as potential interlopers along the line of sight.  

As discussed in \citet{fum14}, our sightlines are selected to ensure that the higher redshift 
LLSs (the natural blocking filters) fully block the quasar light at the wavelengths of our observations. In fact, in most 
cases ($20/32$) the higher redshift absorber is a DLA or a super LLS, and in all 
cases it is a highly optically-thick ($\tau > 2$) LLS. For a typical quasar spectral energy distribution (SED)
$F_{\rm qso} \propto \lambda ^{-1.4}$, and given the redshift distributions of the intervening LLSs, 
the  transmitted flux integrated in the filters used for this imaging survey is $\ll 1\%$ 
of the intrinsic flux for blocking filters with $N_{\rm HI} = 10^{19}~\rm cm^{-2}$, and $< 2\%$ for 
a blocking filter with $N_{\rm HI} = 10^{18}~\rm cm^{-2}$. We therefore exclude leakage from the 
background quasars as an important contaminant in our measurements. This will also  become evident in the 
following discussion, given that only 3 DLAs show emission at the position of the absorber. Of these, two cases are
clearly associated to a neighboring galaxy.

The 10:G11 sightline deserves a particular note. 
From the gallery presented in Figure \ref{fig:smallgallery}, one can
see emission right at the quasar position. For this DLA, a strong LLS at 
$z_{\rm lls}=4.4671$ acts as the blocking filter, but we can only constrain 
the column density to a lower limit of $N_{\rm HI} > 10^{17.8}~\rm cm^{-2}$. 
Damping wings are not readily visible in the spectrum, but 
we cannot trivially set a robust upper limit, as this is a proximate system.  
For the observed quasar flux density 
$F_{\rm qso} \sim 4.7 \times 10^{-29}\rm erg~s^{-1} cm^{-2} Hz^{-1}$ at $\sim 5000$\AA, the upper limit
of the quasar flux leakage within the imaging filter is $F_{\rm qso} < 1.7 \times 10^{-30}\rm erg~s^{-1} cm^{-2} Hz^{-1}$.
As shown in Table \ref{tabinsfr}, the flux detected against this quasar is below this limit, and therefore  we cannot
exclude leakage as a source of contamination for this sightline.

Throughout this work, fluxes ($f_\nu$) are converted into luminosity ($L_\nu$) and SFR ($\dot\psi$) 
with the calibration \citep[cf.][]{fum10b}
\begin{equation} 
\dot\psi = 7.91\times10^{-29} 4\pi D^2_{\rm lum}(z) f_\nu K_{\rm corr}(z) \phi_{\rm igm}(z)
\end{equation}
where  $D_{\rm lum}$ is the luminosity distance to the DLA, $K_{\rm corr}$ 
is the K-correction, and $\phi_{\rm igm}$ accounts for the fact that our observations probe flux 
blueward of 1215\AA\ in the DLA rest-frame and thus should be corrected for absorption in
the IGM. We emphasize that the quoted SFRs bear significant uncertainty, 
especially given that we probe the FUV SED at $\lesssim 1200$\AA\ in the DLA rest-frame. 
At these wavelengths, UV fluxes are 
not an optimal tracer of recent star formation due to both a flattening of the SED around 
1000-1100\AA, and the onset of strong absorption lines 
\citep[e.g.][]{lei02}. Nevertheless, to first approximation, a simple power law  
$f_\lambda \propto \lambda^{-\beta}$ with $\beta = 2$ 
can be assumed to model the FUV continuum between $\sim 1000-1500$\AA\ 
to within a factor of two. This error, albeit large, is comparable to the scatter 
in the FUV slopes from galaxy to galaxy and it is also comparable to the scatter in the model SEDs 
for different input metallicities and star formation histories. A power-law index 
$\beta = 2$ further allows us to adopt standard calibrations at 1500\AA, and to trivially compute K-corrections. 

We estimate the IGM correction for each DLA by combining the appropriate 
filter transmission curve with the mean IGM transmission, computed following standard procedures 
\citep[e.g.][]{mad95}, but using the updated calculations by \citet{ino14}. 
We also account for the fact that the UV flux is fully absorbed at wavelengths blueward of the DLA 
Lyman limit. Additionally, absorption from molecular gas and metals are considered negligible 
\citep{not08,bec13,jor14b}. As discussed above, we do not correct for intrinsic dust absorption, which we argued is 
modest in DLAs. The final limits on the SFRs are listed in Table \ref{tabinsfr} where formal $2\sigma$ upper 
limits are marked by the '$<$' sign to distinguish them from the few instances in which positive flux is detected. 
The listed errors account only for the uncertainty in the flux, without systematic errors on the SFR calibration.  

\begin{figure}
\includegraphics[scale=0.33,angle=90]{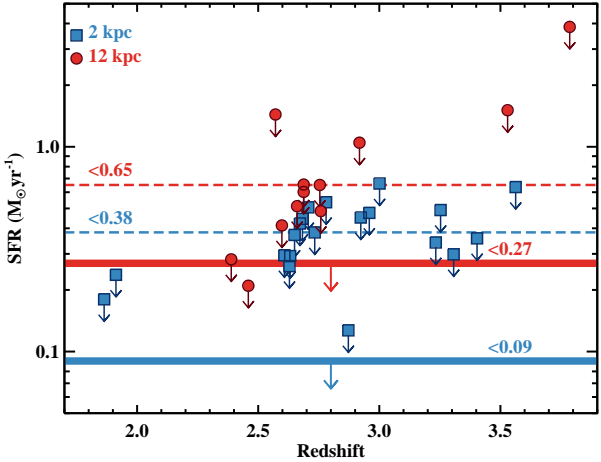}
\caption{Upper limits 
on the {\it in-situ} SFRs of DLAs in two apertures of 
2 and 12 kpc (blue squares and red circles), computed respectively for the HST sample and for
the ground-based sample. Limits are at $2\sigma$ for non-detections, and correspond to the measured flux 
for the three detections (see text for details). Filled symbols are for individual DLAs, while the dashed lines 
indicate the median upper limits of the two distributions. Upper limits measured in the composite images are 
shown instead with thick blue and red lines.}\label{fig:sfrind}
\end{figure}

The measured upper limits on the {\it in-situ} SFRs of individual DLAs are shown in 
Figure \ref{fig:sfrind}, both for the HST sample (blue squares) and for the ground-based 
sample (red circles). Assuming the median upper limits as a typical reference value
(dashed lines in the figure), we conclude from the HST imaging that DLAs do not 
form stars at rates $\gtrsim 0.38~$\sfr\ within compact regions of $\sim 2~\rm kpc$ in size. Similarly, the 
ground-based sample indicates that typical DLAs do not form stars in extended regions
of $\sim 12~\rm kpc$ in size with SFRs $\gtrsim 0.65~$\sfr.
However, differently from the HST sample in which we do not have positive detections, 
for the ground-based sample we detect flux at more than $2\sigma$ C.L. in $3/12$ apertures. 
As noted above, we cannot exclude that the flux detected against the DLA 10:G11
is free from quasar contamination, thus conservatively we conclude that 
$\le 9\%$ of the DLAs form stars with a rate above our sensitivity limits, and that virtually 
no DLAs form {\it in-situ} stars at these rates within compact clumps. 
Given the small sample size, these percentages bear significant uncertainty, 
but they offer the first quantitative description of the local star forming 
properties of DLA gas in the high redshift Universe from FUV tracers.

As for the positive detections, without spectroscopic redshifts, we do not know if these sources 
are genuine DLA galaxies. However, for a given magnitude $m'$ and projected distance $r'$ from the DLA, 
we can compute the expected number $N'$ of interlopers with $m \le m'$ and $r\le r'$ by using the 
observed galaxy number counts in deep $U$ and $B$ band images. Using the published values by \citet{kas04} and 
\citet{gra09}, we find $N' \sim 0.09$, $N' \sim 0.01$, and $N' \sim 0.0002$ for 3:G3, 5:G5, and 10:G11 respectively.
This translates into a low probability of finding random projected galaxies at the separations of 
5:G5 and 10:G11. Conversely, the probability that 3:G3 is an interloper is not completely negligible.
Keeping the caveat of possible leakage in mind for 10:G11, 
it is plausible that some of these detections, and especially 5:G5, are physically associated to the 
DLAs (or lower redshift absorbers). Spectroscopic follow-up of these candidates is now necessary, although this represents a non-trivial task given their faint magnitudes ($\gtrsim 25.5$).

\begin{figure}
\includegraphics[scale=0.5]{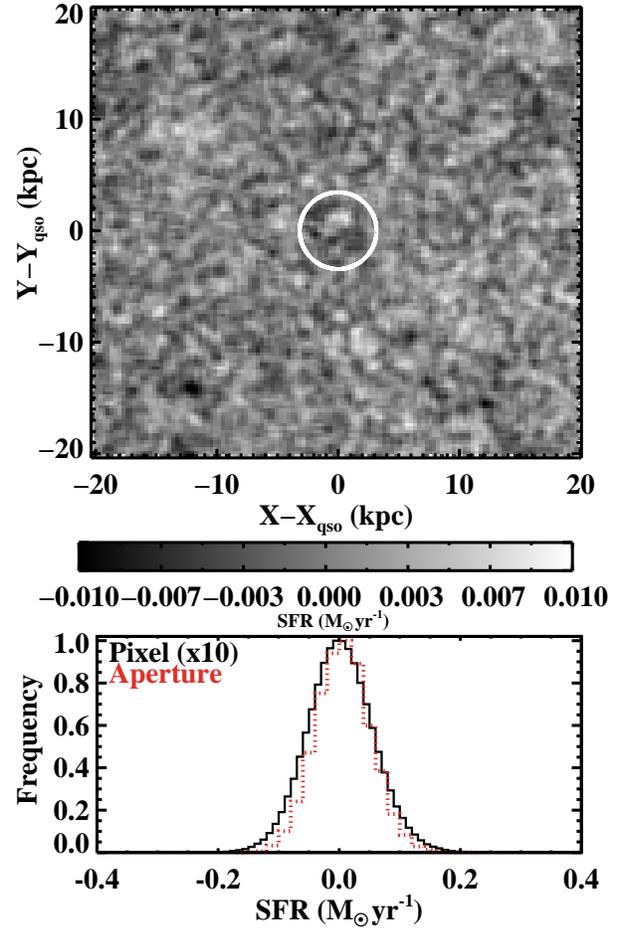}
\caption{{\it Top.} Spatial distribution of the median SFRs in a box of 40 kpc centered at the DLA 
position for the 20 quasar fields imaged with HST WFC3 UVIS. This map has been smoothed with a kernel of 3 pixels 
for visualization purposes. The circle represents the 2 kpc aperture used for the analysis. {\it Bottom.} 
The black solid line shows the histogram of the SFRs measured in individual pixels with size of 0.32 kpc. 
The SFRs that enter this histogram have been multiplied by a factor of 10 to improve the visibility of the
full width. The red dotted lines show instead the distributions of SFRs measured within 10000 random apertures, 
the sizes of which have been matched to the aperture shown in the top panel.}\label{fig:stackhst}
\end{figure}

\begin{figure}
\includegraphics[scale=0.5]{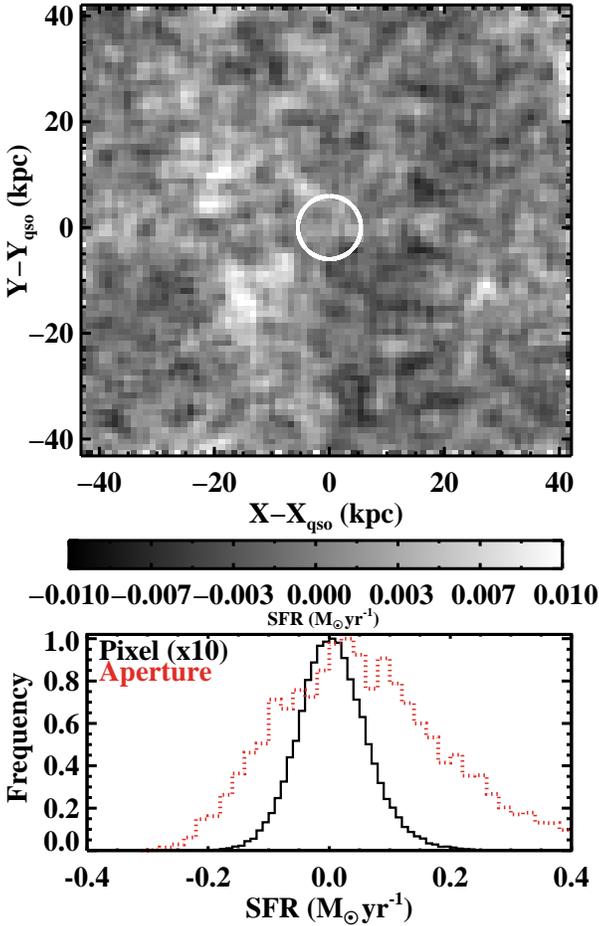}
\caption{Same as Figure \ref{fig:stackhst}, but for the median of the 12 fields imaged 
from the ground. Here we show a region of 80 kpc centered at the DLA position. The image 
has been smoothed with a kernel of 2 pixel. Each pixel in the composite image has a 
size of 1.1 kpc.}\label{fig:stackgrn}
\end{figure}

\subsubsection{SFRs in composite images}\label{ins:stack}

To probe even fainter SFRs, we generate two composite images by separately combining  the 20 HST images and
the 12 ground-based images. These stacks are generated after converting each image to physical units
by applying the SFR calibration discussed in Section \ref{ins:direct}. We also map 
angular separations into physical distances. In this way, we remove the redshift dependence from 
field to field, which is shown in Figure \ref{fig:sfrind}. To create the composite 
images, we then redistribute the SFRs measured in individual pixels onto a grid of pixel size 0.32 kpc and
1.1 kpc, for the HST and ground-based samples respectively. This pixel size, close to 
the native image pixel size at the median redshift $z=2.7$, ensures a fairly uniform mapping of the original 
images in the final stack, with $\sim 19$ and $\sim 11$ independent pixels entering each pixel in the final composite 
for the HST and ground-based samples, respectively. Given that we are not interested in accurate
image reconstruction, we adopt a simple shift-and-add algorithm that preserves the total 
flux. 

When producing stacks, a choice has to be made on what statistics should be used in combining 
images. Given the small sample size, especially for the ground-based sample, the mean 
stack shows a significant numbers of positive detections across the field, due to the large number of galaxies 
which lie in proximity to the quasars in these deep images. The corresponding flux distribution as a function 
of distance from the DLAs encodes interesting information regarding the clustering 
of sources near the targeted absorbers. 
We will consider the above in Section \ref{dlagal}.
In this section, we are interested in characterizing the typical {\it in situ} star forming 
properties of the DLA gas. We hence choose to use the median statistic as a better estimator
of the SFRs of typical DLAs, given that it is less sensitive to the bright pixels
associated with the few positive detections within the regions of interest.
During our analysis, we empirically reconstruct the noise properties of the two stacks
rather than propagating the error map from individual images.
The two median images are shown in Figure \ref{fig:stackhst} and Figure \ref{fig:stackgrn}. 
As in the previous analysis, we compute the SFRs within two apertures of 2 and 12 kpc diameters
to constrain the {\it  in-situ} star formation in compact and extended regions using the 
HST and ground-based imaging, respectively. We also measure SFRs within a larger aperture 
of 4 kpc in the HST image to test for the presence of more extended emission in this sub-sample. 

A proper assessment of the error is critical to establish whether a low flux level is present in the stacked images. 
The pixel standard deviation of the composite images, shown in the bottom panel of Figure \ref{fig:stackhst} and Figure 
\ref{fig:stackgrn}, is well-behaved, implying that our median images are free from large-scale gradients. Furthermore, this 
pixel standard deviation, computed empirically across the image, is consistent with the typical pixel variance recovered from 
a bootstrap technique with 500 iterations. Therefore, we can use the pixel variance as our estimator of the noise in the 
composite image. To further account for correlated noise, we estimate the flux standard deviation for the science apertures
by measuring the standard deviation of the recovered SFRs in 10000 apertures which 
we randomly position inside the composite images far from the quasar location. 

For the HST sample, the distribution of SFRs in these 
randomly-distributed apertures is well approximated by a Gaussian with standard deviation $0.045~$\sfr. 
Therefore, the SFR $\dot\psi = 0.011~$\sfr, measured in the 2 kpc aperture, is 
consistent with the range of SFRs detected in random apertures. We hence place a 
$2\sigma$ upper limit of $\dot\psi < 0.090$\sfr\ on the median {\it in situ} SFR 
of the DLAs of the HST sample. For the ground-based sample, instead, the 
distribution of SFRs measured in random apertures is more skewed towards positive values. 
This is because, differently from the HST sample, both the smaller sample size and the higher 
density of detected sources bias the median of each pixel towards positive values 
(in a correlated fashion). 
Nevertheless, this distribution can still be approximated by a Gaussian with standard deviation 
$0.14~$\sfr. It follows that the measured SFR $\dot\psi = 0.13~$\sfr\ within a 12 kpc aperture at the 
quasar position is statistically consistent with random noise, and we set a 2$\sigma$ upper limit of 
$\dot\psi< 0.27~$\sfr\ for the DLAs of the ground-based sample. 
These limits are summarized in Table \ref{tabinsfr} and Figure \ref{fig:sfrind}.

Finally, no flux is detected in the larger 4 kpc aperture in the HST composite, and we obtain a 
2$\sigma$ upper limit of $\dot\psi< 0.23~$\sfr. Low surface brightness flux from even more extended 
sources would be buried under the noise, which increases as a function of the number of pixels 
within the aperture. For comparison, \citet{red09} report $M_{*}=-20.97 \pm 0.14$ at 1700\AA\ as the characteristic luminosity for star-forming galaxies between $2.7 \le z < 3.4$, and thus our limits at $\le 1100$\AA\ translate into SFRs which are typical for galaxies with $\le 0.01L_*$ for the HST composite and $\le 0.03L_*$ for the ground-based composite. 

Before proceeding further, we validate the adopted procedure  by combining 
the HST images after artificially inserting at the DLA position in each image 
a source with size 1.6 kpc and SFR that is distributed as a Gaussian with median $\dot \psi = 0.11~$\sfr. After processing these fake images through the same analysis pipeline, we recover an SFR 
$\dot\psi = 0.119 \pm 0.045~$\sfr\ in the composite image. Similarly, when we insert a source
with size 5.5 kpc and SFR distributed around a median $\dot \psi = 0.25~$\sfr\ in each image of the ground-based sample, we recover a median $\dot \psi =0.38\pm 0.14~$\sfr\ from the stack. 
Thus, in both cases we recover the input SFRs within the errors.

\subsubsection{Comparison with previous work}

Both the analysis of individual systems and the study of the two composite images corroborate a picture in which 
DLAs do not form stars {\it in situ} with rates $\dot\psi \gtrsim 0.5~$\sfr. With the 
higher resolution HST images, we also exclude the presence of compact star forming regions with 
SFRs as low as $\dot\psi \sim 0.1~$\sfr\ at the position of the quasar. 
Keeping in mind that the conversion between the observed UV fluxes and the
SFRs are subject to a substantial degree of uncertainty, we can put these limits in the context of previous attempts
to establish the SFRs of DLAs. 

By stacking spectra of DLAs with $\log N_{\rm HI} \ge 20.62$  and examining the residual flux in the 
Ly$\alpha$ trough, \citet{rah10} set a $2\sigma$ upper limit on the Ly$\alpha$ emission from DLA gas at 
$<2.0\times 10^{-18}~\rm erg~s^{-1}~cm^{-2}$, which corresponds to $< 0.02L_*$ at $z\sim 2.8$ for their assumed 
\lya\ luminosity function, or $\dot\psi <0.8~\rm M_\odot~yr^{-1}$ given their adopted conversion 
between \lya\ flux and SFR. From a re-analysis of the same data, \citet{rau11} concluded instead 
that flux is positively detected to $(5.35\pm 1.97) \times 10^{-18}~\rm erg~s^{-1}~cm^{-2}$ when restricting the 
analysis to positive velocities compared to the systemic DLA redshift, as expected for resonant lines. However, 
the detection of \lya\ in this composite spectrum is complicated by subtle systematic errors and it is only marginal. 

In fact, more recent work by \citet{cai14} sets an even more stringent limit on the \lya\ emission from a composite 
of $\sim$2,000 DLAs to $<0.01L_*$ at $z\sim 2.6$. Emission is detected when restricting to the highest column density
DLAs with $\log N_{\rm HI} > 5\times 10^{21}~\rm cm^{-2}$ \citep{not14}, which are however 
a small subset of the general DLA population. Considering the many uncertainties at play in the two different measurements, 
we conclude that both the analysis of \lya\ in composite spectra and our direct measurement of the FUV flux are 
in agreement and rule out the presence of appreciable {\it in-situ} star formation in DLA gas to comparable limits.

\begin{figure}
\includegraphics[scale=0.33,angle=90]{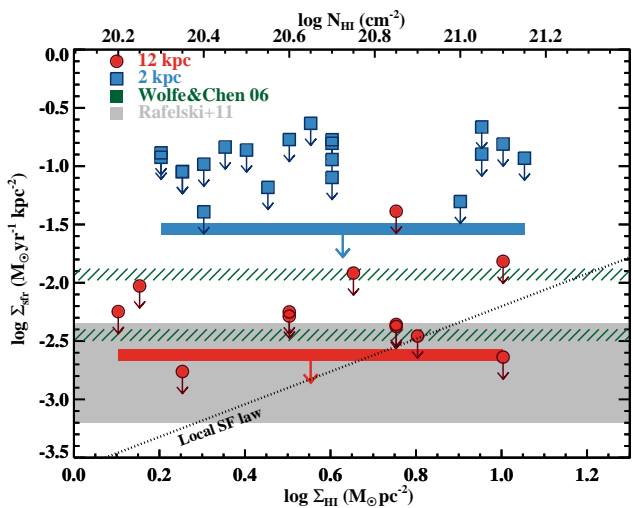}
\caption{Limits on the SFR surface densities $\Sigma_{\rm sfr}$ of DLAs  from the HST sample (blue) and  the 
ground-based sample (red). 
Limits for individual systems are shown as squares and circles for the two samples, respectively. Limits derived from the 
composite images are shown as horizontal bars. Also shown with a grey shaded region is 
the range of SFR surface densities detected in the outskirts of LBGs by \citet{raf11} and, with green dashed bands, the 
limiting surface densities probed by \citet{wol06} for two different kernel sizes 
($0.25''$ for the upper value, and $1.0''$ for the lower value). The dotted 
black line shows where the extrapolation of the local SF law would lie at the lower column densities
that are typical for DLAs.}\label{fig:sflaw}
\end{figure}

\subsection{The star formation law of DLAs}\label{ins:sflaw}

Having established direct limits on the {\it in-situ} SFRs in a representative sample of DLAs, 
we now turn to the relationship between the neutral gas observed in absorption and the rate with 
which new stars form, as probed in emission. Empirically, the link between gas and star formation is described 
via a star formation (SF) law. One common parametrization of the SF law, 
known as the Kennicutt-Schmidt law \citep{sch59,ken98}, is in the
form $\Sigma_{\rm sfr} = K (\Sigma_{\rm gas}/\Sigma_{\rm gas,0})^\beta$, where $\Sigma_{\rm sfr}$ and 
$\Sigma_{\rm gas}$ are the SFR and gas surface densities, with constants $K$, $\Sigma_{\rm gas,0}$, and $\beta$.
While comparing $\Sigma_{\rm sfr}$ and $\Sigma_{\rm gas}$ in our observations, we should bear in mind 
that pencil beam surveys intrinsically probe scales of few parsecs, while our SFR surface density extends to scales of few 
kpc. Therefore, this does not allow us to formally define in quantitative terms an SF law, which is known to be scale-dependent in the nearby Universe \citep[e.g.][]{sch10}. For this reason, and because the measured upper limits do not 
allow us to constrain the SF law parameters in detail, here we only present a general discussion of the link 
between $\Sigma_{\rm sfr}$ and $\Sigma_{\rm gas}$, also in comparison with previous studies. 

Figure \ref{fig:sflaw} shows that, once recast as SFR surface densities, our limits 
become $\log \Sigma_{\rm sfr} \lesssim -0.90~$\Ssfr\ in the compact regions probed by the HST imaging, 
and $\log \Sigma_{\rm sfr} \lesssim -2.25~$\Ssfr\ in the more extended regions probed by the ground-based imaging. 
The composite images provide even tighter constraints on the median SFR surface densities of the DLA 
population: $\log \Sigma_{\rm sfr} < -1.54~$\Ssfr\ for the HST sample, and $\log \Sigma_{\rm sfr} < -2.62~$\Ssfr\
for the ground-based sample. Once compared to the \HI\ column densities, we see that DLAs could, in principle, 
consume the available neutral gas reservoir quite rapidly, despite their modest {\it in-situ} SFRs \citep[cf.][]{pro09}.

For a range of DLA column densities that brackets 
our sample ($N_{\rm HI}= 10^{20.3}-10^{21.3}~\rm cm^{-2}$), 
we can infer lower limits on the depletion time scales, which lie in the range $>0.06-0.56$ Gyr 
given the median SFR limits of the HST sample. Similarly, 
for the ground-based sample, the depletion time scales lie in the range $>0.66-6.66$ Gyr. 
These values are simply a reflection of the fact that for typical DLAs, our limits lie for the most part 
above the extrapolation of the local SF law considering for example the original fit by \citet{ken98} with 
$\beta = 1.40 \pm 0.15$, $K_{\rm disk}=(2.5\pm 0.7)\times 10^{-4}~$\Ssfr, and $\Sigma_{\rm gas,0}=1~\rm M_\odot~pc^{-2}$. 
As shown by Figure \ref{fig:sflaw}, only the SFR limits for the highest column density DLAs in the 
ground-based sample are comparable to the extrapolation of the local SF law. Therefore, 
deeper limits, by $1-2$ orders of magnitude, would be required to probe the {\it in-situ} SFRs if the DLA gas had 
to form stars according to the extrapolation of the local SF law. This finding is in line with previous 
studies, as discussed in Section \ref{prinsit}.

Given the recent progress in understanding the phenomenology and the physics of the 
SF law, we should also consider from a theoretical point of view whether some star formation, if any, is 
to be expected in DLA gas with column densities $N_{\rm HI}= 10^{20.3}-10^{21.5}~\rm cm^{-2}$. Modern studies 
\citep[e.g.][]{won02,fum08,big08,fum09,big10,gne10,ost10,fel11,kru12,glo12,ler13}
have shown that new stars preferentially form in regions of higher gas surface density and metallicity,
where gas can cool efficiently. Conversely, star formation is reduced or completely suppressed in regions 
of low metallicity and at low column densities. More quantitatively, both simple numerical models or 
detailed hydrodynamic calculations \citep[e.g.][]{kru09b,gne10} reveal the existence of a 
threshold in $\Sigma_{\rm gas,t}$ below which $\Sigma_{\rm sfr}$ sharply drops.
At the low metallicity of DLA gas with $Z\sim 0.1Z_\odot$, $\Sigma_{\rm gas,t}\sim 10~\rm M_\odot~pc^{-2}$.

While the exact relationships between metallicity, molecular gas, and star formation are still being 
investigated \citep[e.g.][]{glo12,kru12b}, high {\it in-situ} SFRs in DLAs should not be common  
given that these systems lack a favorable environment for star formation \citep[cf.][]{kru09}. 
Deeper imaging surveys that use the same strategy adopted here have the potential of 
directly constraining the physics of star formation in the lower column density and lower 
metallicity regions common to DLAs, offering a direct test of these models. 
This will be possible with future 30m telescopes, but it is also within the reach 
of current facilities, provided one carefully selects  targets in narrow redshift ranges so as 
to maximize the FUV emission within the imaging filters. 

\subsubsection{Comparison with previous work}\label{prinsit}

A few studies have already investigated the star-formation law in DLAs, using UV as 
a tracer for star formation \citep[e.g.][]{wol06,raf11}. However, because of the bright quasar glare, these analyses rely on 
indirect or statistical methods to link the \HI\ column density seen in absorption to the UV emission. 
The first of these studies is the work by \citet{wol06}, who searched for extended low-surface brightness 
galaxies with sizes between $2-31$ kpc in the Hubble Ultra Deep Field (HUDF). This search yielded
for the most part non-detections, which were used by the authors to constrain the comoving SFR densities 
of DLAs at $z=2.5-3.5$. Comparing these limits to the  SFR densities expected if DLAs were to 
form stars according to an SF law, \citet{wol06} concluded that star formation in DLAs 
proceeds with an efficiency\footnote{In agreement with previous DLA studies, in this section,
we define the constant $K$ in the star-formation law as efficiency, noting that modern investigations on 
the SF law prefer the definition of depletion time for the inverse of this quantity.} 
of less than 10\% compared to nearby galaxies. More recently, the study by \citet{raf11} revisited the problem of the 
SF law in DLAs, under the assumption that these absorbers arise instead in the outskirts of compact LBGs. By stacking 
a sample of $\sim 50$ LBGs at $z\sim 3$ in the HUDF, \citet{raf11} detected FUV emission to radii of 
$\sim 6-8$ kpc from the LBG centres. After connecting the observed SFR densities to the 
column density distribution of DLAs, they concluded that the SFR efficiency of this gas is a factor 
of $10-50$ lower than what one would predict from the local SF law.

Figure \ref{fig:sflaw} offers a comparison between our survey and these previous results, 
after rescaling all SFRs to the calibration assumed in this work. 
Our composite images probe SFR surface densities which are comparable
to the ones probed by \citet{wol06} for both compact and more extended regions. However, the original work of 
\citet{wol06} does not provide direct limits on the SFR surface density of DLAs, but only limits on the SFR comoving 
density of the DLA population. Thus, our analysis complements and extends the results of their investigation, as it 
provides direct observational evidence that in fact DLAs do not form stars {\it in-situ} within  
regions of appreciable UV surface brightness. If these limits are extrapolated to all DLAs up to a column density
of \NHI$\sim 10^{22}~\rm cm^{-2}$, our observations naturally explain the lack (or the scarce number) 
of low-surface brightness sources in the HUDF, reinforcing the conclusions of the statistical 
study of \citet{wol06}.

Compared instead to the analysis of \citet{raf11}, limits from our composite HST image 
are not informative, given the superior depth of the HUDF and the larger sample size of LBGs used by these 
authors to create their stack. Our deep ground-based imaging, however, places more interesting constraints on the 
SFRs in an extended low-surface brightness component, similar to the one probed around LBGs by \citet{raf11}. 
At first, one may conclude that the limit on the median SFR surface density which our study recovers  
is in tension with the surface brightness detected in the LBG composite by \citet{raf11}. 
However, this is not the case for two reasons. 

First, \citet{raf11} produced a stack of compact LBGs, centering at 
the position of the peak surface brightness. Thus, for an underlying radial surface brightness profile, 
faint flux levels are coherently stacked when moving away from the LBG center. Conversely, in our stack, we coadd 
images centering at the DLA position, thus off-centre compared to the core of a putative nearby LBG. 
This has interesting consequences for the expected number of LBGs within a given projected distance from the quasar,
and we will return to this point in the next section. For the purpose of this discussion, we note instead that, 
because DLAs can arise from many impact parameters, different surface brightness levels are not
coadded coherently in our stack. This 
geometric effect causes an intrinsic dilution of the signal, and thus our limits are not necessarily at odds with the 
measurement of \citet{raf11}.

Secondly, and perhaps most importantly, according to the model proposed by \citet{raf11}, the detected emission arises 
from column densities $N_{\rm HI} \gtrsim 10^{21}~\rm cm^{-2}$, which are not typical of 
the DLAs in our sample. For this reason, and given the above discussion about the SF law, the measurement of \citet{raf11} 
applies to only a small fraction of the DLA population. Intriguingly, a similar trend has been seen in studies of 
\lya\ emission in the DLA trough, where flux is detected in composite spectra of the highest and most rare column 
density DLAs \citep{not14}, but not when considering the more typical lower column density systems \citep{cai14}
included in our study. Our measurement and the previous discussion on the SF law thus highlights how the results 
of \citet{raf11} should not be simply extrapolated to the more general DLA population, outside of the 
column density interval that has been explored by our study.

\begin{figure}
\includegraphics[scale=0.33,angle=90]{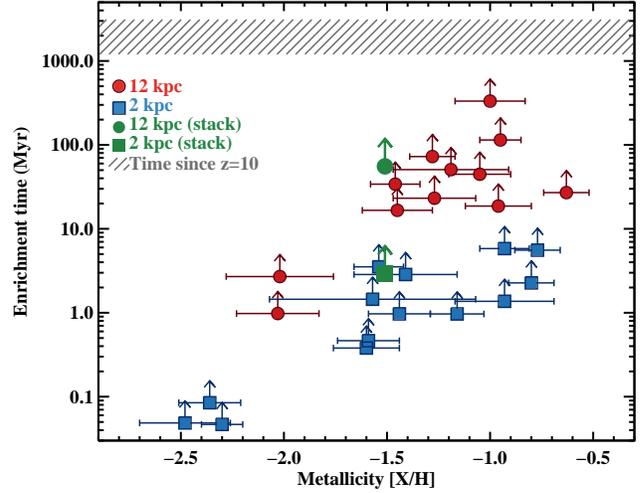}
\caption{Lower limits on the time that is needed for the {\it in-situ} star formation to enrich the
DLA gas to the observed levels. As in previous figures, we show limits derived in 2 kpc apertures 
from HST imaging with blue squares, and limits derived in 12 kpc apertures from ground-based imaging 
with red circles. The green circle and the green square refer to the limits obtained using the 
composite images in each of the two samples. The dark grey shaded region indicates the elapsed 
cosmic time between $z=10$ and the DLA redshifts, which we assume as the maximum time available 
for the enrichment. This calculation relies on the assumption that the \NHI\ measured in 
absorption is representative of the \HI\ column density within the adopted apertures.}\label{fig:metal}
\end{figure}

\subsection{Implications for metal production}\label{ins:metal}

By leveraging the new vantage point offered by our observations, we can 
compare directly the observed {\it in-situ} star formation of DLAs in emission
to the observed metal content of DLAs in absorption. During this part of
the analysis, we restrict to the sub-sample of 26 DLAs for which we have high quality spectroscopy to 
compute the neutral gas phase metallicity, as discussed in \citet{fum14}. 
As emphasized for the previous analysis of the SF law, the absorption spectroscopy and the available
imaging do not constrain the physical properties of DLAs on the same scales. 
For instance, the recent analysis by \citet{kan14} shows how the 
\HI\ column densities measured on the smallest scales probed in absorption differ 
from the values measured when smoothing on much larger spatial scales. 
Thus, the results presented in this section rely on the crude assumption that the measured 
column densities of hydrogen and metals are, to first order, representative of the typical DLA 
gas on scales of $\sim 2-12$ kpc. 
 
To link the limits on the SFRs and the measurements of metallicity, we 
compare the mass in metals within DLAs ($M_{\rm Z,obs}$) to the upper limits on the 
mass in metals produced by {\it in-situ} star formation ($M_{\rm Z,sfr}$), given the constraints 
from our imaging observations. With this comparison, we infer lower limits on 
the enrichment time $\Delta t_{\rm Z}$, a quantity which in turn offers a simple metric for 
whether the {\it in-situ} SFR alone is potentially sufficient to enrich DLAs, or whether external 
contributions are needed. Specifically, we compute the mass in metals within DLAs as 
$M_{\rm Z,obs} = 10^{\rm [X/H]} Z_\odot (m_{\rm p} N_{\rm HI} \pi r^2)$, where $\rm [X/H]$ is
the metallicity measured in absorption and $Z_\odot = 0.0181$ is the metallicity of the Sun
\citep{asp09}. In this equation, the term in parenthesis translates the \HI\ column density to the total
hydrogen mass, assuming a cylinder of diameter $r=2$ and $r=12$ kpc for the HST and ground-based 
imaging, respectively. 
Given the negligible ionization of DLAs, the neutral gas phase is assumed to trace the bulk of the 
mass in the system. For this reason, our calculation is restricted to metals within DLAs, 
without considering metals in an ionized phase, which may nevertheless be physically associated to these 
absorbers \citep{fox07,leh14}.

Next, we compute the expected mass in metals related to star formation
as $M_{\rm Z,sfr} = y_{\rm Z} \dot\psi \Delta t_{\rm Z}$, where $y_{\rm Z}$ is 
the metal yield, weighted according to a given initial mass function. 
Several choices for $y_{\rm Z}$ are available in the 
literature \citep[see, e.g., the discussion in][]{pee14}, with modern 
values ranging between $\sim 0.02-0.04$. In the following, we assume $y_{\rm Z} = 1/42$ \citep{mad96}, 
a common choice for such studies. Finally, comparing the inferred lower limits on
$M_{\rm Z,sfr}$ with the values of $M_{\rm Z,obs}$, we derive lower limits on the enrichment 
time, as shown in Figure \ref{fig:metal}.

Having only upper limits on the SFRs, we cannot set stringent constraints on the history of 
the metal enrichment of DLAs. It is however interesting to note that the amount of metals locked
in the DLAs is fairly modest once integrated over the 2 kpc area used for the analysis of the HST 
sample. Thus, if the DLAs of our sample have SFRs just below the detection limit of the 
HST imaging (i.e. $\sim 0.1-0.6~$\sfr), it would be straightforward to achieve 
the observed enrichment levels within $\sim 0.05-10~\rm Myr$.
For the ground-based sample, integrating over the larger (12 kpc) apertures, 
$\sim 1-300$ Myr would be needed to achieve the observed enrichment levels, if the 
SFRs were at the detection threshold ($\sim 0.2-2~$\sfr).

Given that the available time for star formation to enrich 
DLA gas is $\sim 2~\rm Gyr$ if stars form as early as $z\sim 10$, SFRs as low as
$\dot\psi = 2\times 10^{-4}$\sfr, 500 times lower than the limits derived in 
our HST composite image, can still satisfy the observed metallicity in compact DLAs. A similar result 
holds for more extended DLAs, if star formation proceeds with a rate of $\dot\psi = 9\times 10^{-3}$\sfr, 
30 times less than the limits of our ground-based composite image. 
We therefore conclude that the metals produced by low levels of {\it in-situ} star formation 
are in principle sufficient to account for the modest metal content of DLAs. However, if in fact no 
star formation occurs at the lower column densities common in DLAs, then external sources of 
metals are required to raise the mean DLA metallicity to $\sim 0.1Z_\odot$, which is significantly 
above the metal content of the IGM \citep{sch03,sim11}. Again, deeper versions of this experiment 
have the potential of better constraining the mechanisms through which the densest regions 
of the high-redshift Universe are enriched.

\subsubsection{Comparison with previous work}

Previous work to constrain the enrichment histories of DLAs has compared 
two statistical quantities, i.e. the cosmic density of metals produced by star 
forming galaxies ($\rho_{\rm Z,sfr}$) and the amount of metals that reside in DLA gas at any given 
time ($\rho_{\rm Z,dla}$). The consensus is that DLAs at $z\sim 2-3$ contain only a small fraction 
of the metals produced in star-forming galaxies \citep[e.g.][]{pet99,bou07,raf14}, with the latest estimates 
suggesting  $\rho_{\rm Z,dla} \sim 0.01 \rho_{\rm Z,sfr}$ at $z\sim 2.3$ \citep{raf14}. Albeit 
with substantial uncertainties, most notably not accounting for a potential contribution 
from an enriched and dust-obscured DLA population, it clearly appears 
that DLAs are not enriched by all the metals produced in galaxies. 
A similar ``missing metal problem'' was found by \citet{wol03}, who compared $\rho_{\rm Z,dla}$
with  a new estimate of $\rho_{\rm Z,sfr}$ in DLAs (and not LBGs) derived from the 
cooling rates inferred from \CIIs\ absorption lines (see also Section \ref{sec:cool}). 

However, the comparison between $\rho_{\rm Z,dla}$ and $\rho_{\rm Z,sfr}$ from indirect constraints 
on the {\it in-situ} SFRs in DLAs has also yielded contrasting results, depending 
on the different assumptions on the nature of DLAs which were made by different authors. 
In particular, combining their inferred star formation efficiency for 
extended low-surface brightness DLAs with the results of numerical simulations,
\citet{wol06} suggested that metals may be underproduced in DLAs compared to the observed values.
When accounting instead for the yields associated with the extended UV emission 
detected in the outskirts of LBGs, \citet{raf11} achieved instead a closure of the 
metal budget in DLAs.

The analysis presented in Figure \ref{fig:metal} is consistent with the statistical
arguments at the origin of the missing metal problem. Given the lack of appreciable local star formation, 
as pointed out by our observations, one can understand why DLAs contain fewer metals than those produced in LBGs, 
which form stars at much higher rates. Furthermore, as shown in Figure \ref{fig:metal}, DLAs can be enriched 
to the observed values even by low levels of star formation, either {\it in-situ} or, 
if $\dot\psi\ll 10^{-2}-10^{-4}~$\sfr, by modest episodes of star formation occurring in the immediate 
surroundings \citep[e.g.][]{raf11}. Conversely, at the limits of our observations or those 
by \citet{wol06}, it seems unlikely that we are facing a problem of underproduction of metals in DLA gas.

\subsection{Implications for cooling rates}\label{sec:cool}

The \CIIs$\lambda1335.7$ absorption line arises from the $^2P_{3/2}$ level in the ground 
state of ionized carbon (C$^{+}$), and thus provides a tool to measure the column density of 
C$^{+}$ ions in the $j=3/2$ state. This is of much interest as these ions give rise 
to the $\rm [CII]158\mu$m transition ($^2P_{3/2} \rightarrow ^2P_{1/2}$), 
which is believed to be the major coolant of the neutral ISM \citep[e.g.][]{wol95}. 
The strength of the \CIIs$\lambda1335.7$ absorption can hence be used to measure the cooling 
rate ($\ell_{\rm c}$) of DLA gas \citep{pot79,wol03a}, and, under the assumption
of thermal balance, to infer the corresponding heating rate.
Leveraging this idea, \citet{wol03a,wol03}
have developed a formalism to quantify the intensity of the radiation field that is responsible for the 
photoelectric heating of the DLA gas, and hence for the corresponding SFR surface density. 

\begin{figure}
\includegraphics[scale=0.33,angle=90]{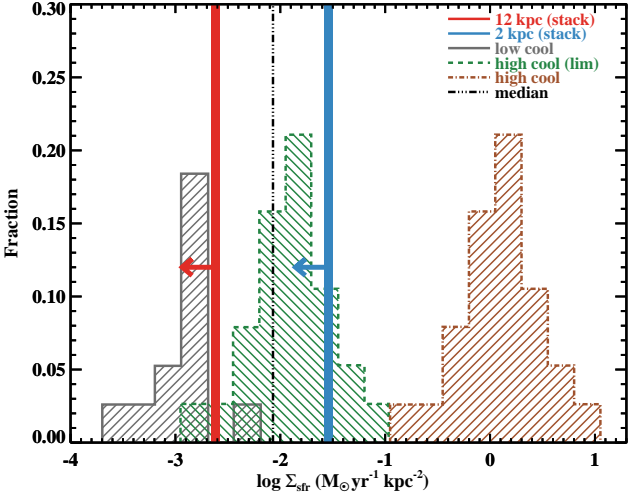}
\caption{A comparison between the SFR surface densities inferred by \citet{wol08} from \CIIs\ 
  absorption (histograms) and the limits inferred from our composite images (vertical lines). 
  From left to right, the grey solid, green dashed, and brown dot-dashed histograms refer 
  respectively to low cool DLAs, lower limits for the high cool DLAs, and values corrected 
  for the filling factor of the star forming regions for the high cool DLAs. The median 
  of the observed sample (low cool and limit on high cool) is also shown with a black triple dotted 
  line. The \citet{wol08} values have been corrected by a factor 1.58 to account for the different SFR 
  calibration used in our study. The vertical lines are for the SFR surface density limits measured in 
  the 2 kpc aperture in the HST composite image (blue line to the right) and in the 12 kpc aperture 
  in the ground-based composite image (red line to the left).}\label{fig:cool}
\end{figure}

Applying this formalism to a sample of 38 DLAs for which $\Sigma_{\rm sfr}$ could be inferred,  
\citet{wol08} noted a bimodal distribution in the inferred star formation surface densities, 
which reflects the bimodality in the DLA cooling rates. Measurements of $\ell_{\rm c}$ show two
different populations, one of ``high-cool'' systems and one of ``low-cool'' systems, 
separated at around $\ell_{\rm c}\sim 10^{-27}~\rm erg~s^{-1}~H^{-1}$\citep[see also][]{nee13}.
While a low cooling rate can be balanced by the heating rate from the metagalactic UV background 
alone, \citet{wol08} concluded in their study that an additional heating input is generally required by their model
to describe these observations.  These authors therefore computed the SFR surface densities needed to satisfy 
the measured cooling rates, under the assumptions detailed in \citet{wol03a,wol03}. 
These distributions are shown in Figure \ref{fig:cool} for both the low-cool 
(solid gray histogram) and the high-cool (green dashed histogram) DLAs.

The ability to measure the UV surface brightness in the same DLAs with detected \CII*\ absorption would 
offer a direct test of this model, in turn providing interesting constraints on the cooling and 
heating mechanisms at work in DLA gas. Unfortunately, our survey is designed to target $z\sim 2.7$ 
DLAs in $z\gtrsim 3.5$ quasars, for which the \CII*\ absorption is redshifted inside the \lya\ forest,
hampering a reliable determination of the cooling rates in individual systems. 
However, we can proceed with a statistical approach, by comparing in Figure \ref{fig:cool} the limits 
on the {\it in-situ} SFR surface densities measured in the composite HST and ground-based images, 
with the surface densities inferred by \citet{wol08} from the observed \CII*\ cooling rates. 

Considering the low-cool DLAs at first, we see that even the stringent limits derived
from the composite images do not provide interesting constraints. A factor of 10 deeper survey 
is needed to probe the nature of this sub-population.
Considering the high-cool DLAs instead, the null detection in our ground-based composite image
appears at odds with the $\Sigma_{\rm sfr}$ implied by the high-cool DLA populations
for extended galaxies. As shown in Figure \ref{fig:cool}, the median $\Sigma_{\rm sfr}$ in the 
ground-based sample is about 0.5 dex lower than the median of the $\Sigma_{\rm sfr}$ inferred from the 
cooling rates when including both low-cool and high-cool, under the assumption that 
our sample is not biased towards one of the two populations (but see below).
In their work, \citet{wol08} reached a similar conclusion on how a population of star forming 
DLAs which satisfy the $\Sigma_{\rm sfr}$ required by the high-cool population would violate the SFR surface 
density limits imposed by the \citet{wol06} analysis. This is not surprisingly, given the previous discussion 
on the limits shown in Figure \ref{fig:sflaw}. 

As noted by \citet{wol08}, this tension can however be alleviated if 
compact LBGs in proximity to, but not fully embedded with, the DLAs act as heating sources. Because of simple 
geometrical arguments related to the $r^{-2}$ dependence of the flux, however, the inferred SFR 
surface densities for the high-cool population should be corrected upward by the relative filling factor of 
star forming regions and DLA gas. Once this correction is applied, the distribution of $\Sigma_{\rm sfr}$ for the high-cool 
DLAs shifts to higher values, as shown in Figure \ref{fig:cool} with a brown dot-dashed histogram.
However, our observations constrain this possibility: if the heating 
arises from compact LBGs embedded in DLA gas, these sources must be located at distances 
$>1-6~\rm kpc$ (i.e. the radii of the adopted apertures throughout our analysis) from the DLA position, 
as indicated by the lack of emission at $\Sigma_{\rm sfr} \gtrsim 0.1$\Ssfr\ in the HST and 
ground-based samples (Figure \ref{fig:sflaw}). 
It is however possible that compact LBGs with high $\Sigma_{\rm sfr}$ lie
at larger impact parameters, a hypothesis we will explicitly test for in Section \ref{dlagalcool}
when studying the distribution of star-forming galaxies in proximity to DLAs. 

Before proceeding, we caution the reader that the basis for the comparison presented in this 
section is that, if the bimodal distribution of the cooling rates and 
the inferred $\Sigma_{\rm sfr}$ are typical of DLAs, than comparable SFRs should be detected in our sample which is 
representative of the DLA population. However, it should be noted that the cooling rates of DLAs correlate with 
metallicity, with the majority of high-cool systems having $\rm [X/H] > -1.2$ \citep{wol08}. The distribution of 
metallicity of DLAs in our study peaks instead around $\rm [X/H] \sim -1.5$  \citep[see figure 14 in][]{fum14}. 
Furthermore, throughout this analysis, we are assuming an equal split between the low-cool and high-cool 
population, following \citet{wol08}. However, the estimate of \citet{wol08} is based on a small sample of DLAs 
and may not reflect the actual distribution in the general DLA population. Our non-detection could therefore be simply
explained by a larger percentage of low-cool DLAs, for which we do not expect UV flux in excess to our 
sensitivity levels. For these reasons, and also because of the small number of systems in our sample, any tension 
between our observations and the inferred $\Sigma_{\rm sfr}$ from \CIIs\ absorption should be regarded only as 
tentative at this point. Future work in larger samples, and in particular a direct comparison of the 
observed versus inferred SFR surface densities in individual systems, is needed to confirm this apparent tension.

\section{The connection between DLA\lowercase{s} and galaxies}\label{dlagal}

In Section \ref{insitu}, our discussion focused entirely on the emission properties of DLA gas, 
directly at the position where spectroscopy in absorption reveals the presence of neutral hydrogen 
with column densities 
above $N_{\rm HI} \ge 10^{20.3}~\rm cm^{-2}$. In this section, we expand our study beyond 
the properties of the {\it in-situ} SFRs of DLA gas, by considering associations between DLAs and 
star-forming galaxies at all impact parameters $b_{\rm dla}$, which we define here
as the projected quasar-galaxy separation.

Given that our images probe only the rest-frame FUV emission, our analysis will only consider 
galaxies with appreciable unobscured SFRs, typically $\dot\psi \gtrsim 1-2~$\sfr. 
Following early studies that employed color selection techniques to identify a population of 
star-forming galaxies at $z > 2.5$ \citep[e.g.][]{ste96,mad96,gia02},  the term LBG has become a synonym for star-forming galaxy in the literature. Throughout this work, we follow the same convention. 
We clarify however that the term LBG here refers to any star-forming galaxy, regardless of its actual 
SFR. Traditionally, LBG samples included primarily UV-bright galaxies with SFRs of tens to hundreds of \sfr, 
but our study also probes  galaxies with lower SFRs, by a factor of 10 to 100. 
In this discussion we will make a distinction between ``bright'' LBGs and ``faint'' LBGs, or dwarfs. 
With the expression bright LBGs, we will refer to a population with properties similar to the original samples
of spectroscopically-confirmed LBGs, with SFRs $\dot\psi\sim 20-50~$\sfr\ and halo masses $\sim 10^{12}~\rm M_\odot$.
Conversely, with the expression faint LBGs, we will refer instead to a population of star-forming galaxies with SFRs $\dot\psi < 10~$\sfr, down to the sensitivity limits of our survey. For this second population, we will also distinguish between faint LBGs which are isolated, i.e. central to their 
dark matter halos, from faint LBGs which are satellites of brighter LBGs. The reason for this 
distinction will become apparent in the following discussion. 

Finally, we note that in the remainder of this paper, we will be naturally biased against a 
population of star-forming galaxies that is significantly dust obscured (see the discussion in 
Section \ref{dustobs}). In Section \ref{insitu}, 
we argued that the low dust extinction seen along the quasar line of sight rules out 
severe dust obscuration as the cause of the observed lack of appreciable {\it in-situ} star formation. 
However, when considering galaxies at all impact parameters, we do not have any prior 
knowledge of the dust properties. Thus, the conclusions drawn from our following analysis 
apply to a galaxy population with unobscured SFRs comparable to our detection limits. 
Galaxies below this detection limit could in principle be still forming stars
to appreciable intrinsic rates. Unfortunately, with our current data, we cannot assess the importance of dust obscuration in 
searches of DLA galaxies, which remains an open question that future studies at 
infrared and millimetre wavelengths should address.

\begin{figure*}
\includegraphics[scale=0.55]{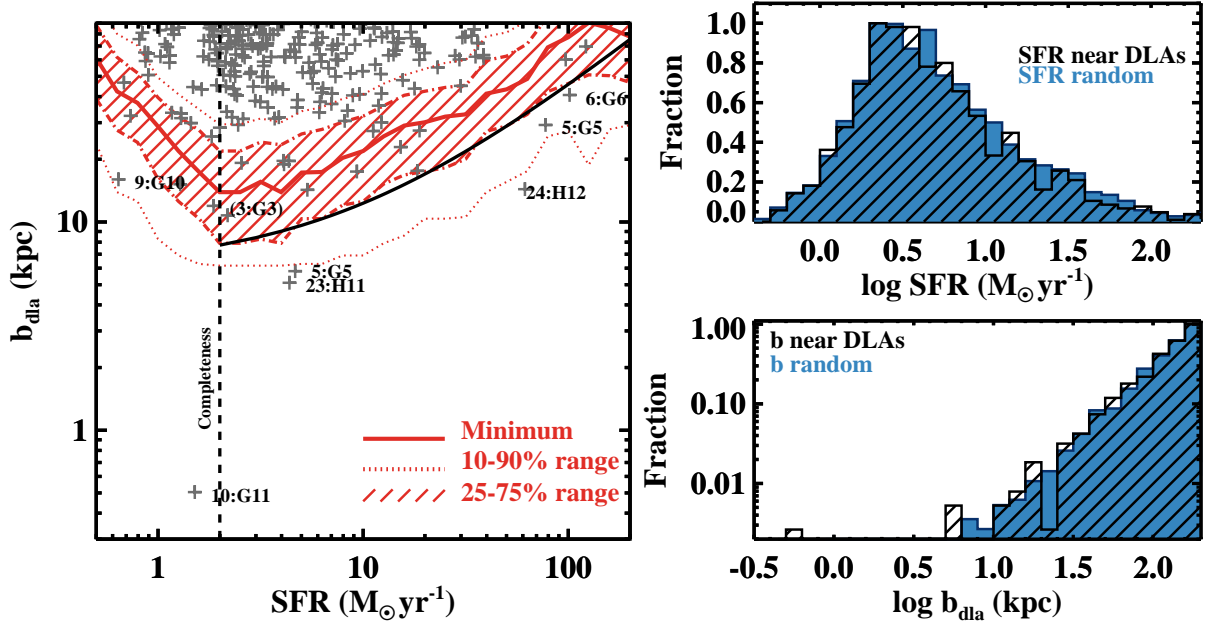}
\caption{{\it Left:} Distribution of impact parameters as a function of SFRs for the galaxies 
  near DLAs. The grey crosses are for galaxies detected near the DLAs, and values 
  are computed at the DLA redshift. Errors on SFRs are not shown 
  to improve visibility, but they range between $5-20\%$ (statistical). The red solid curve 
  represents the median minimum impact parameters computed in 10000 trials by 
  drawing from a ``background'' galaxy population, i.e. random galaxies detected 
  far from the quasar position. The corresponding percentiles of these minimum impact parameters 
  are shown with dotted and dash-dotted curves, respectively for the $10-90\%$ and 
  $25-75\%$ of the distribution.  The black curve shows a numerical approximation for the $25\%$ 
  of the distribution, which holds up to the completeness limit indicated by a vertical dashed line.
  Outlying galaxies (i.e. those lying below the expected
  minimum impact parameters for a random galaxy population) are labeled with the name of their 
  quasar. {\it Right:} Normalized histograms of the SFRs (top) and impact parameters (bottom) 
  for the galaxy population around the DLAs (black dashed histograms) and for the background 
  galaxy population.}\label{fig:bsfr}
\end{figure*}

\subsection{Limits on the impact parameters and SFRs}\label{impactb}

Without a deep and dense redshift survey of the galaxies in the targeted quasar fields,
we cannot establish unique DLA-galaxy associations for all the absorbers. However, 
we continue in the spirit of the previous analysis by studying the distribution 
of SFRs and impact parameters of the identified galaxies to derive statistical constraints 
on the properties of the DLA galaxy population. We start our analysis by focusing on the galaxies 
in proximity to the quasar sightline ($b \lesssim 10-30~\rm kpc$). From a physical 
point of view, associations between DLAs and galaxies at these separations are useful to 
constrain models in which DLAs arise from a gaseous disk or a more amorphous 
structure in the surroundings of an embedded star-forming disk \citep[e.g.][]{fyn08,raf11,dan14}. 
More practically, this is the interval of impact parameters for which our observations become the most
constraining, given the low number of random interlopers predicted within the small projected area subtended by impact parameters of $b \lesssim 30~\rm kpc$.

To derive limits on the SFRs and impact parameters of DLA galaxies, 
we produce for each quasar field $k$ a ranked list of impact parameters with associated SFRs 
$\{b_{\rm dla,i} , \dot\psi_i\}_k$, assuming that each galaxy $i$ is at the DLA redshift. 
For this analysis, we limit our search to $b_{\rm dla} = 200~\rm kpc$, 
which corresponds to $\sim 2$ times the virial radius of a massive LBG at $z\sim 2.7$ 
\citep[e.g.][]{fum14a}. This distance is considered sufficient to encompass both DLAs that arise
from the immediate surroundings of galaxies and systems that originate from gas structures 
(including but not limited to tidal debris or accreting filaments) associated with the halo 
of a galaxy. In this ranked list, the impact parameters can be 
generally considered as lower limits of the true impact parameters because,
if the $i-$th galaxy is not the DLA host, the $i+1$ galaxy will become the next candidate 
with $b_{\rm dla,i+1} > b_{\rm dla,i}$. However, if the $i-$th galaxy is in fact the true DLA galaxy, 
or the DLA galaxy is fainter than our detection limit, then  $b_{\rm dla,j}$ for $j>i$ 
(or even for each $j$) will be distributed as dictated by random galaxy counts.
It follows that, by comparing the observed $\{b_{\rm dla,i} , \dot\psi_i\}_k$ distribution
in the quasar surroundings with the distribution of random galaxies $\{b_{\rm ran,m} , \dot\psi_{\rm m}\}_k$, 
we can identify which galaxies have impact parameters and SFRs which are unusual 
given random galaxy counts, thus constraining the impact parameters and SFRs of the DLA 
galaxy population. 

We present results from this comparison in Figure \ref{fig:bsfr}, where we show the distribution 
of impact parameters and SFRs for all the galaxies detected as a function of the distance from 
the 32 DLAs included in this study. In each quasar field, we also select three apertures of 200 kpc 
in radius, which we place at random positions far from the quasar locations. All the galaxies detected in these 
random apertures are used to characterize $\{b_{\rm ran,m} , \dot\psi_{\rm m}\}_k$, that is 
the SFR and impact parameter distributions of a ``background'' galaxy population given random 
counts of galaxies that are not physically associated with the targeted DLAs. These distributions, 
which are shown in the right panels of Figure \ref{fig:bsfr}, are subject to the same selection effects 
as the galaxy population near the DLAs \citep[see a discussion in ][]{fum14}.
Thus, by empirically measuring the background in proximity to the 
regions of interest with the same selection function, 
we automatically account for the field-by-field variation in the sensitivity limits of 
our survey, the differences arising from the fact that we have adopted multiple filters, and potential 
differences in the large-scale distribution of galaxies in the targeted fields.

The main result of this analysis is already apparent from the histograms in Figure \ref{fig:bsfr}.
The distributions of SFRs and impact parameters for the galaxies close to the quasar sightline 
match the corresponding distribution for 
random galaxies far from the quasars, without an obvious excess that can be 
attributed to the DLA hosts. The only noticeable difference is at 
$b_{\rm dla}<10~\rm kpc$, where few galaxies are detected. 
The same excess is visible also as a function of the SFR, 
as shown in the left panel of Figure \ref{fig:bsfr}. 
Even with these few detections, consistent with our previous discussion on the {\it in-situ} SFR, 
the region $b_{\rm dla}<10~\rm kpc$ is almost completely empty, with only $3/32$ detected 
galaxies ($2/32$ if we interpret the detection in the 10:G11 field as due to quasar leakage). 
From this analysis, we draw an important conclusion: not only do DLAs not form appreciable stars 
{\it in situ}, but also that $\le 7\%$ of the DLAs are associated with galaxies with 
$\dot\psi \ge 2~$\sfr\ and $b_{\rm dla}<10~\rm kpc$. In turn, this implies that the majority of DLAs 
arise from galaxies with $\dot\psi < 2~$\sfr\ and/or at $b_{\rm dla}>10~\rm kpc$.

Figure \ref{fig:bsfr} also offers a convenient empirical way to summarize the effective completeness 
of our survey. One can see that the distribution of impact parameters reaches its maximum density 
around $\dot\psi \sim 2~$\sfr, while it is more sparse at both larger and smaller SFRs. 
At larger SFRs, such a decrease is expected according to the general form of the 
luminosity function, for which the number of sources decreases as the apparent magnitude decreases 
(hence the luminosity increases). The decrease of density for 
increasing apparent magnitudes is instead a result of incompleteness in our flux-limited survey. 
In the following analysis, we will consider $\dot\psi = 2~$\sfr\ as our effective completeness limit, keeping in mind 
that the depth of individual images varies from field to field.
We note that the discrepancy between the completeness limit and the more sensitive 
limits on the {\it in situ} SFR arises from the different nature of the two measurements.
While the {\it in situ} measurements are only limited by pixel variance, the completeness limit
also depends on the procedures used to identify sources in the images. We refer the reader to \citet{fum14} for a 
detailed description of these effects.

Beyond $b_{\rm dla}=10~\rm kpc$, one can instead see an increasingly high number of detections, 
which, as suggested by the distributions of SFRs and impact parameters, are for the most part unrelated to the DLAs. 
But are we seeing any excess compared to a pure background distribution, which we can associate with the intervening 
DLAs? To answer this question more quantitatively, we derive an estimate for the locus of the minimum impact parameter 
as a function of SFR, given a random galaxy population and a reference point in the sky. In other words, we seek for
an expression $b_{\rm min} = b_{\rm min} (\dot\psi)$ which quantifies the 
typical distance to the closest galaxy in the sky, given an SFR and a preferred view point as set
by a quasar. We emphasize that in this calculation $\dot\psi$ is a surrogate for the apparent 
magnitude, and it should not be interpreted as an actual SFR measurement. 

To derive $b_{\rm min}$, we simulate the locus of impact parameters as a function of SFRs using the 
distributions of background galaxies with 10000 mock samples. 
Specifically, for each trial, we draw $N_{\rm ran}$ pairs $\{b_{\rm ran,b} , \dot\psi_b\}_k$ 
from the distributions of impact parameters and SFRs of the background
galaxy population, with  $N_{\rm ran}$ equal to the number of sources detected 
within 200 kpc from the 32 DLAs included in this study. We then define 
$b_{\rm min}(\dot\psi)=\min(b_{\rm ran,b}(\dot\psi))$ in bins of SFR. By repeating this procedure 10000 times, we recover 
a distribution $\{b_{\rm min}\}_t$, the median of which is shown in Figure \ref{fig:bsfr}, together with 
the $0.1,0.25,0.75,0.9$ percentiles. To first approximation,  
$b_{\rm min, 25\%} / {\rm kpc} = (\dot\psi/{\rm M_\odot~yr^{-1}})^{0.8}+6$. 
This parametrization should be generalized to other samples with caution,
being related to the extremes of a distribution.

\begin{figure*}
\begin{tabular}{ccc}
\includegraphics[scale=0.21,angle=0]{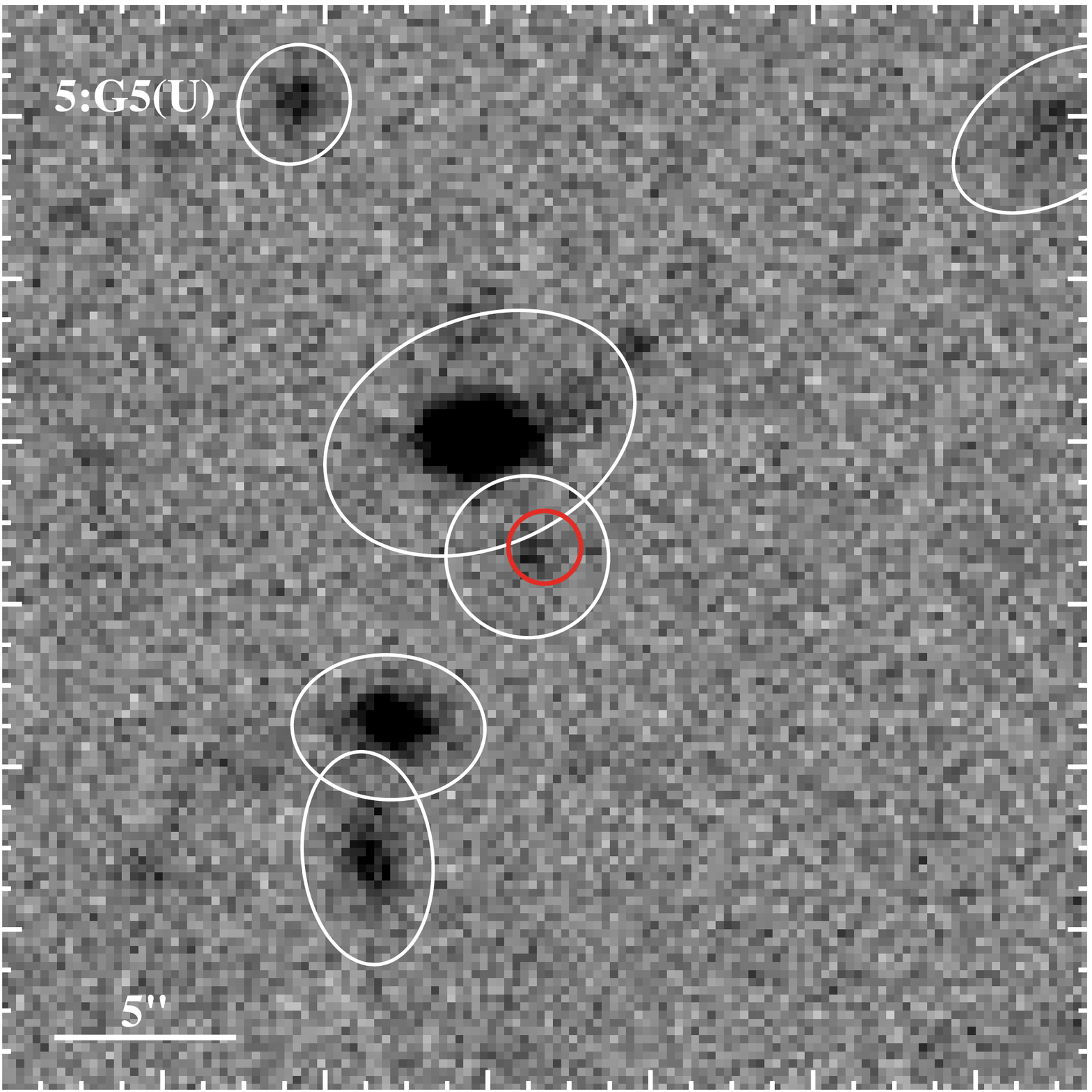}&
\includegraphics[scale=0.21,angle=0]{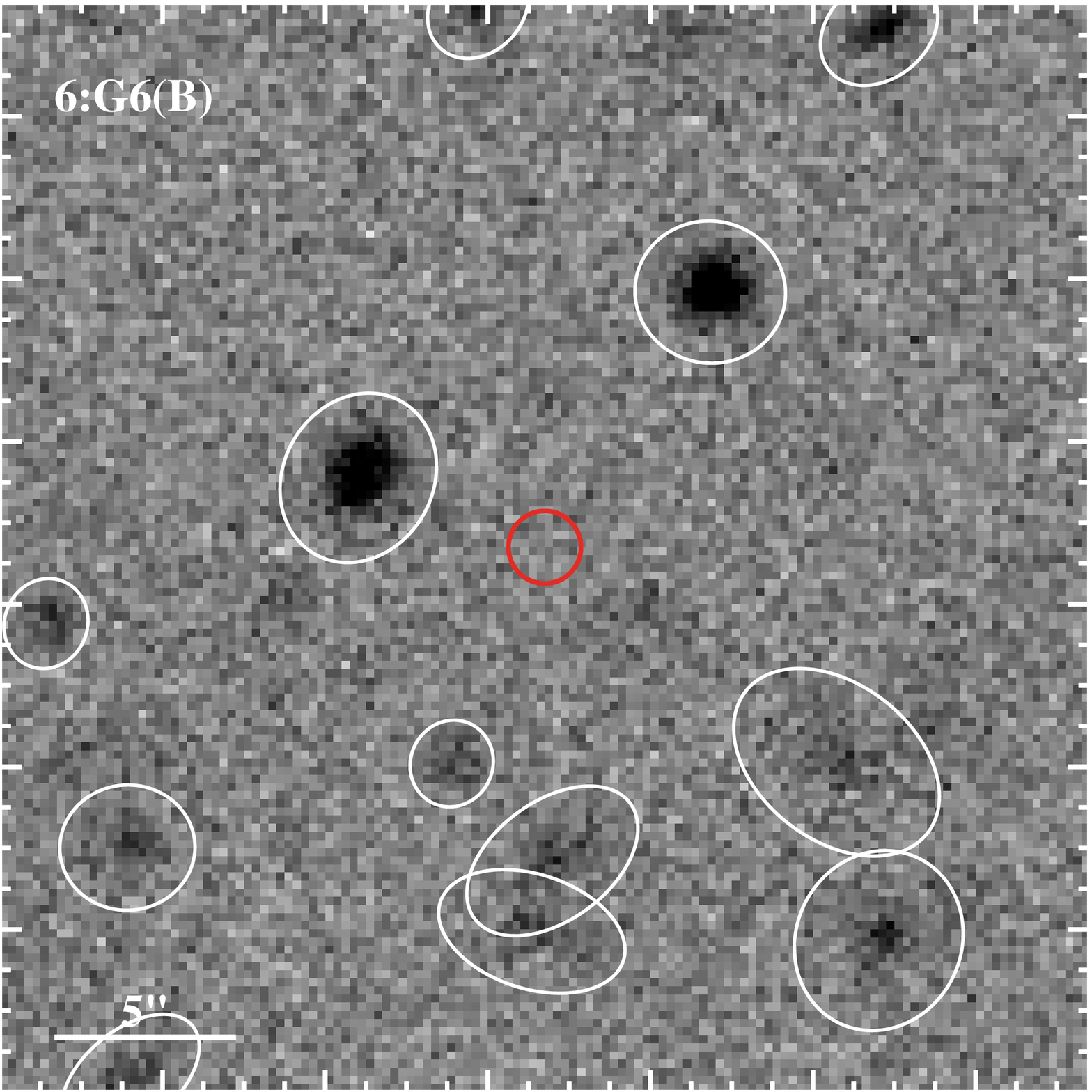}&
\includegraphics[scale=0.21,angle=0]{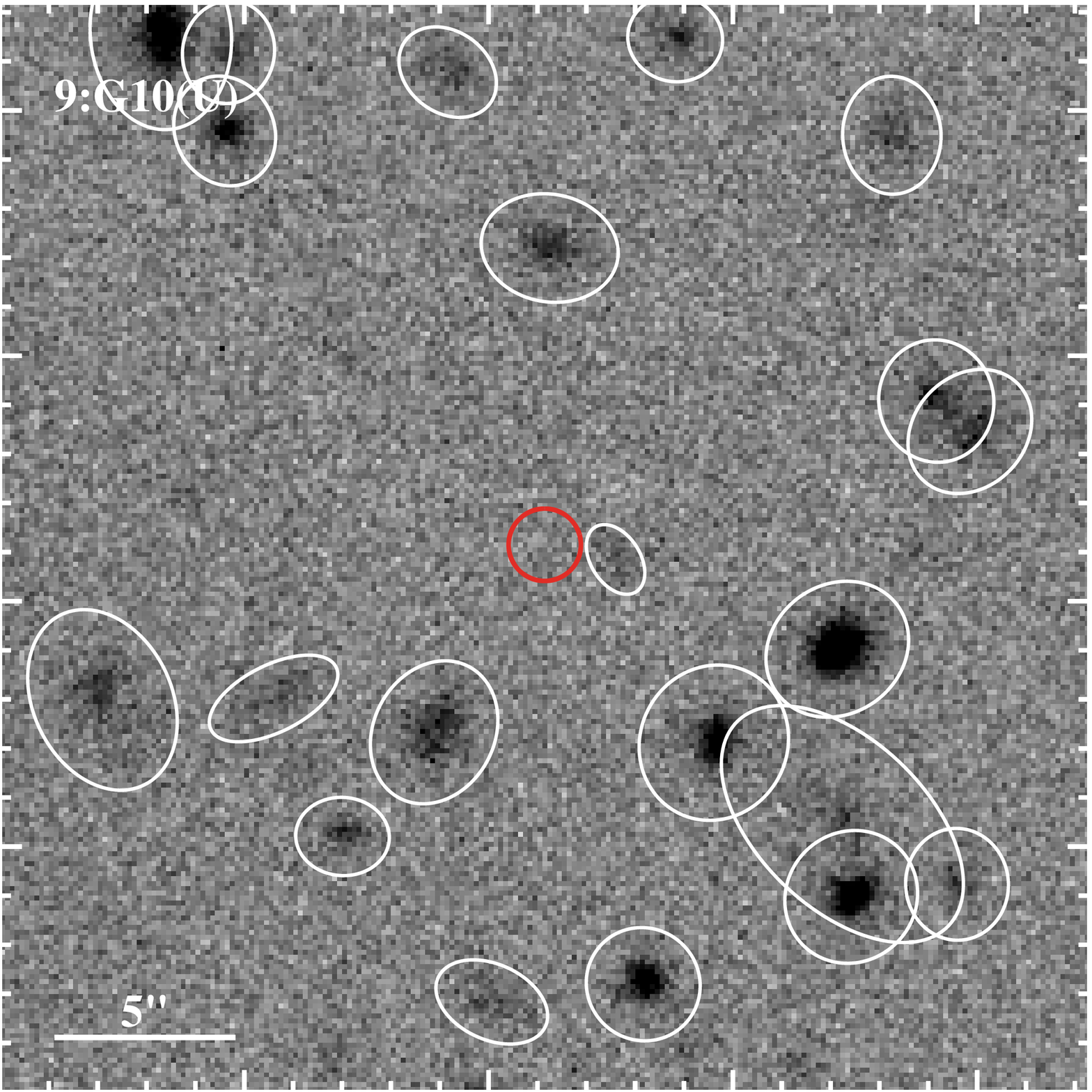}\\
\includegraphics[scale=0.21,angle=0]{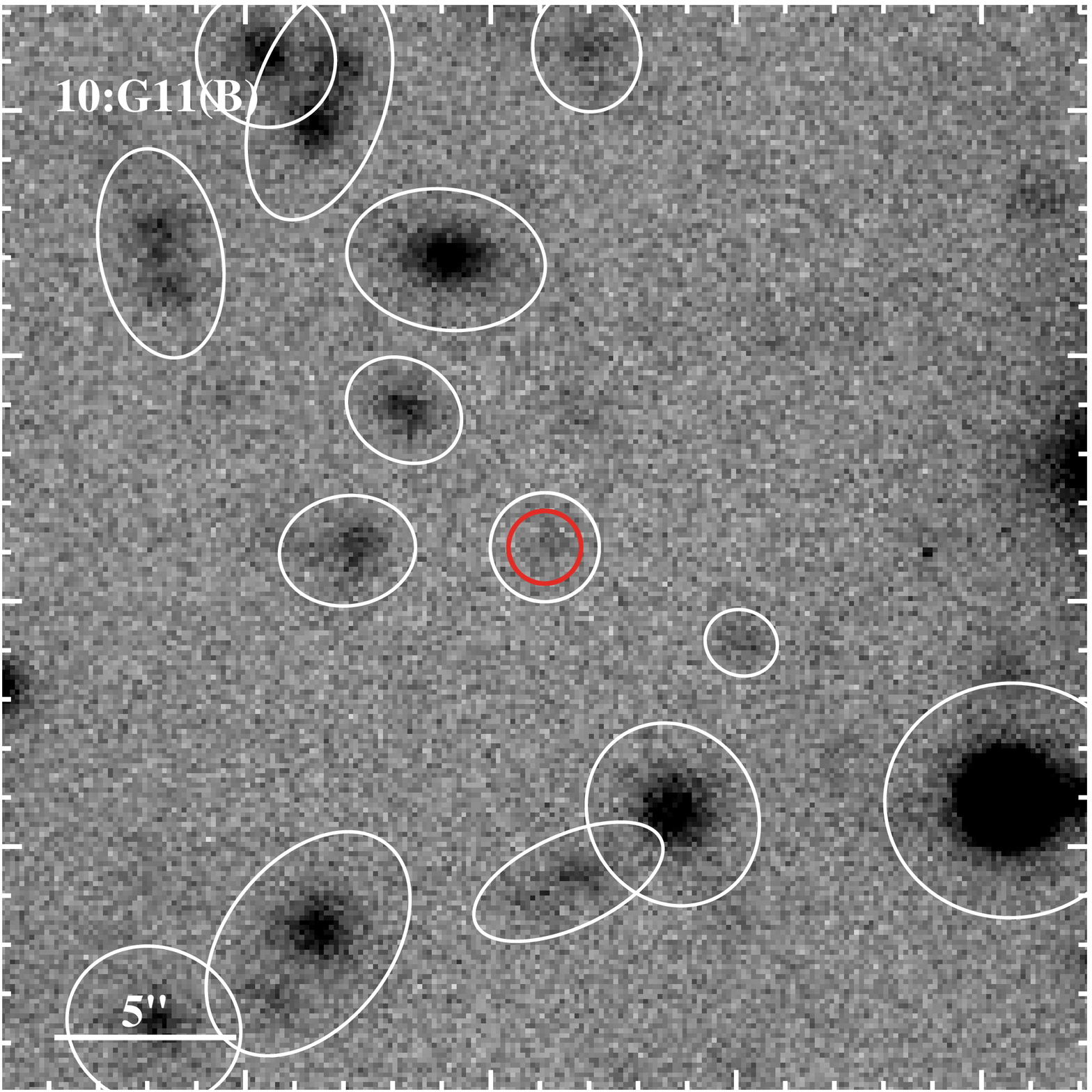}&
\includegraphics[scale=0.21,angle=0]{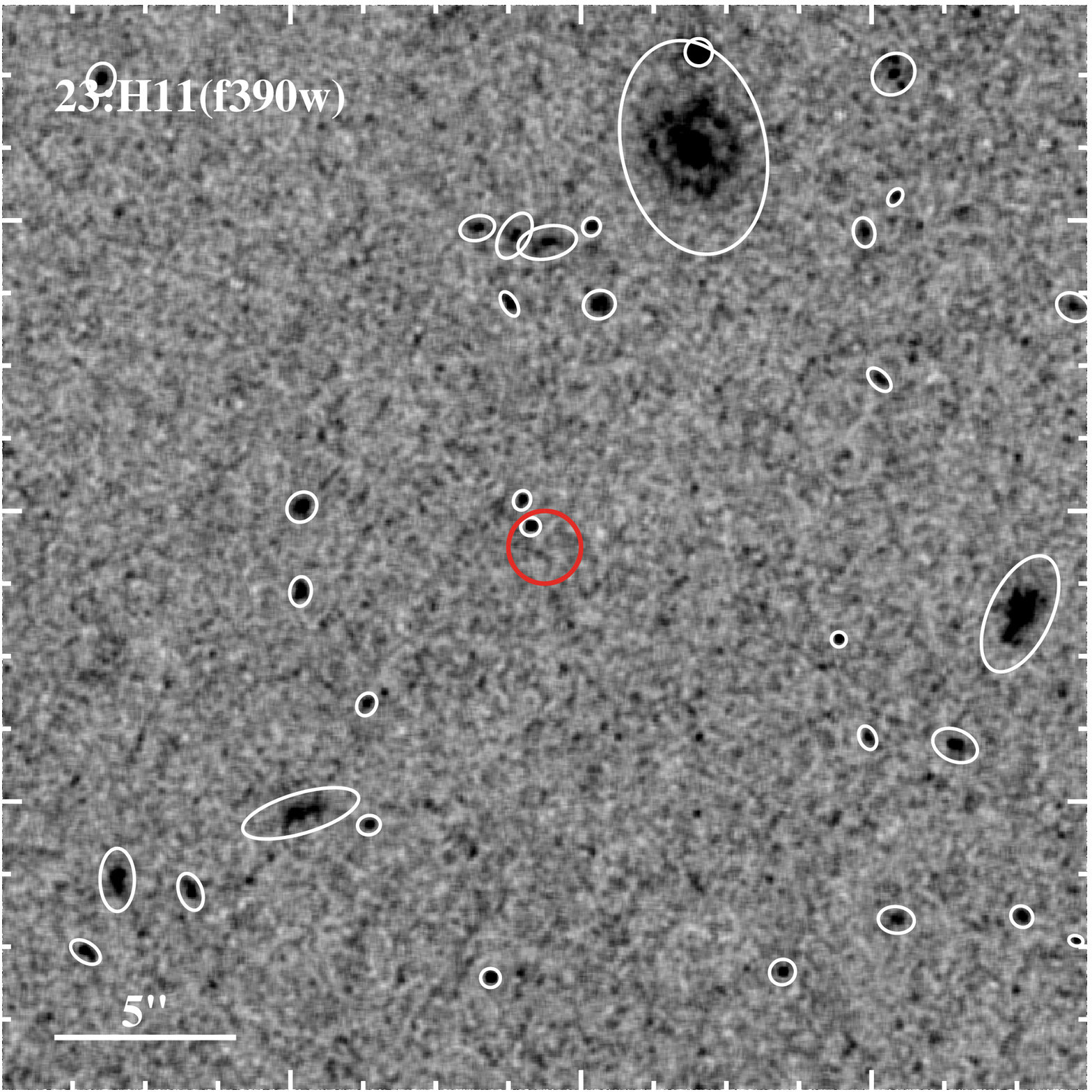}&
\includegraphics[scale=0.21,angle=0]{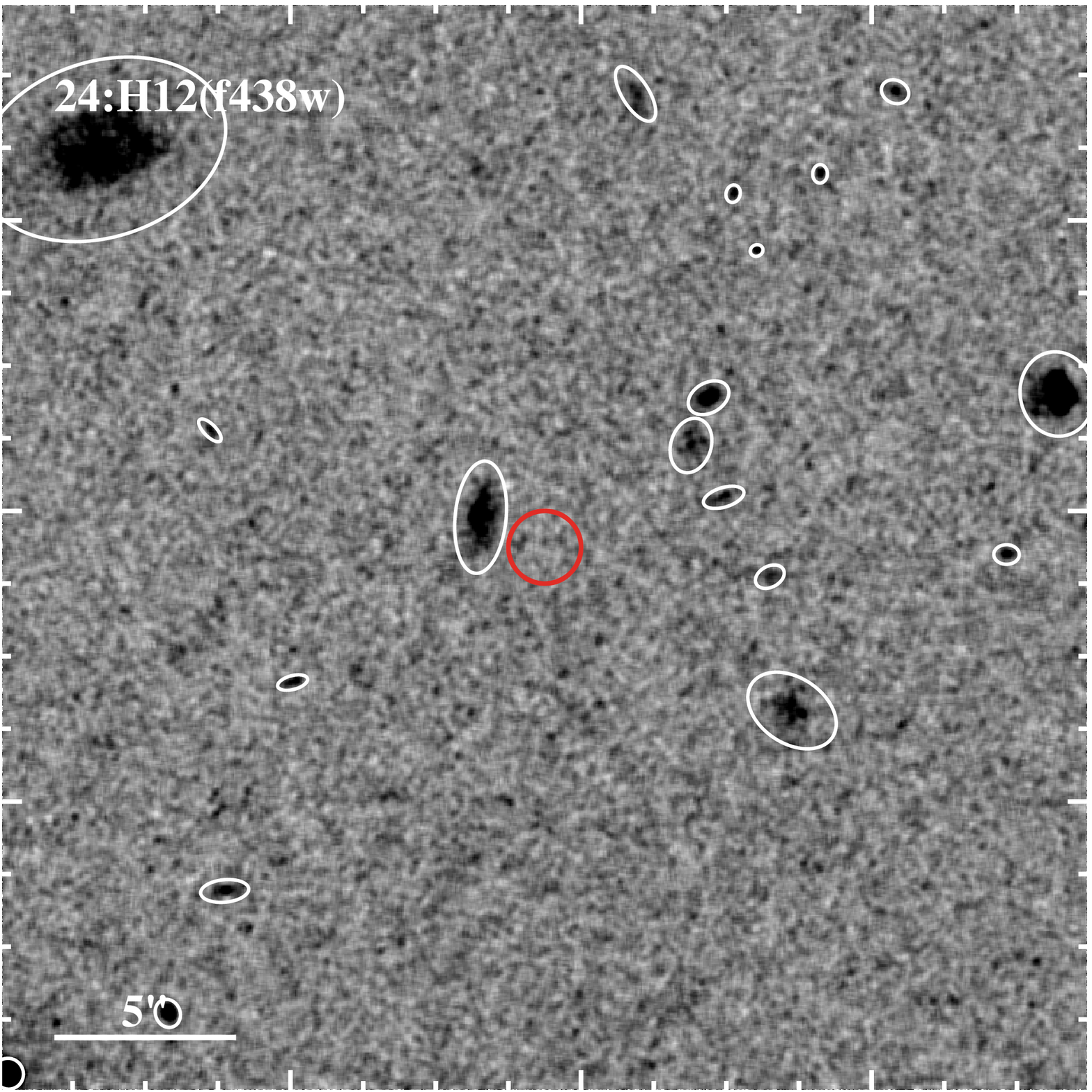}
\end{tabular}
\caption{Images in the bluest available filters of the six quasar 
  fields with candidate DLA galaxies at close impact parameters. 
  Each panel shows a $30''\times 30''$ region centered at the DLA position, which is marked 
  by a red circle. The detected sources are marked with white ellipses, which represent the 
  Kron apertures used for photometry.}\label{fig:smallgallery}
\end{figure*}

With an estimate for the typical $b_{\rm min}$ in a random galaxy sample, we have a metric 
to conclude that there is a statistically significant excess of galaxies at small impact parameters, 
with $4/32$ detections below  $b_{\rm min, 10\%}$, and $6/32$ detections below $b_{\rm min, 25\%}$. 
Images of these 6 quasar fields are shown in Figure \ref{fig:smallgallery}.
Thus, from this analysis, we can conservatively conclude that $\le 13\%$ of the DLAs are associated 
with galaxies with $\dot\psi \ge 2~$\sfr\ and impact parameters $b_{\rm dla} \le b_{\rm min, 25\%}$.
In Figure  \ref{fig:bsfr}, we also label the location of the 
galaxy in the field 3:G3 that gives rise to the positive detection within the
12 kpc aperture centered on the DLA in the previous section. This galaxy lies
close to the typical $b_{\rm min}$ for random galaxies, in line with our previous estimate 
of a $\sim 10\%$ probability to detect a random interloper with comparable magnitude in the 
search area defined by the projected impact parameter. The association of the detected emission 
for this DLA is therefore tentative. 

Nevertheless, we believe that we have detected a handful of 
sources that are likely associated to the intervening DLAs along the targeted sightlines.
We emphasize that, although the target fields were selected due to the presence 
of a DLA and we are hence ``biased'' towards the association of any detected galaxy
with the DLA, we cannot formally rule out associations with lower-redshift absorption
line systems (e.g. MgII or CIV absorbers) which are present in these spectra.
This is especially true given the statistical nature of this argument and 
the small number of positive detections. Thus, the $6/32$ identifications should be 
regarded as quite conservative upper limits. 

Finally, we note that the host galaxies of the higher redshift LLSs and DLAs which act as 
blocking filters should not be detectable below their Lyman limit, unless they are characterized by an unusually high 
escape fraction of ionizing photons. Thus, the presence of higher redshift blocking filters 
does not affect our estimates. In passing, we point out that the search for dropout galaxies in proximity to these 
quasars will offer interesting candidates for the association with the higher redshift DLAs and LLSs in these sightlines. 
At the same time, the search for DLA hosts at $b_{\rm dla} > 10~\rm kpc$ through a statistical analysis like the one 
here presented does not have to be restricted to sightlines with double DLAs, and it can be easily performed in larger 
samples. We defer these searches to future work.

\begin{figure}
\includegraphics[scale=0.33,angle=90]{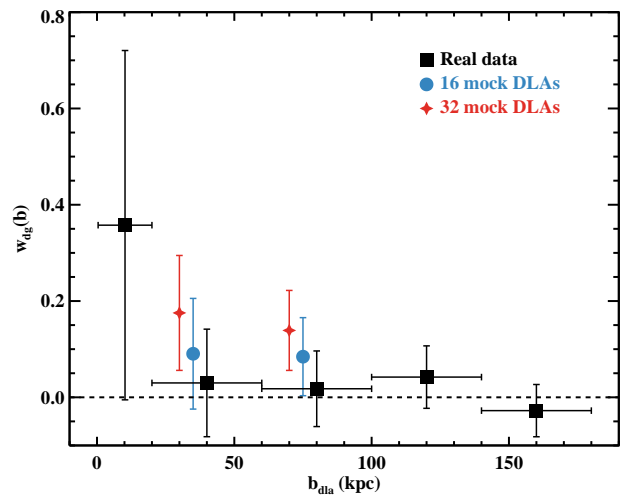}
\caption{The projected DLA-galaxy correlation function as measured around the 32 DLAs included in this study 
  within bins of 40 kpc (except the innermost bin of 20 kpc). The measurement obtained with real data is shown with 
  black squares, while blue circles and red stars show the projected correlation function after injecting 
  16 and 32 artificial DLA hosts with random impact parameters in the range $b_{\rm dla}=20-100~\rm kpc$. For these two 
  cases, we show only the two relevant bins, as the other values remain unchanged. The central values have also been 
  offset by 5 and 10 kpc for visualization purposes.}\label{fig:chidla}
\end{figure}

\subsection{The galaxy-absorber correlation function}

In section \ref{impactb}, we focused our attention on the closest impact parameters, in the range 
$b_{\rm dla} \lesssim  30~\rm kpc$. We now extend our analysis to larger impact parameters to investigate 
whether there is an excess of galaxies near DLAs compared to a background 
galaxy population. To address this question, we measure the projected DLA-galaxy correlation function, 
by comparing the observed number of galaxies at a given separation from the targeted DLAs to the 
expected number from a random galaxy population \citep[see, e.g.,][]{coo06}. Again, for this 
calculation, we consider a circular aperture of 200 kpc around each DLA.

\begin{table*}
\caption{Properties of intervening DLAs with confirmed galaxy associations.}\label{known} 
\centering
\begin{tabular}{l c c r r c r r c c l}
\hline
\hline
Name&$z_{\rm qso}$&$z_{\rm dla}$&$b_{\rm as}$&$b_{\rm p}$&$\log$\NHI       &\multicolumn{2}{c}{SFR}                  &[X/H]& Ref.\\
    &             &             &(")         &(kpc)      &(cm$^{-2}$)&\multicolumn{2}{c}{(M$_\odot$ yr$^{-1}$)}&    & \\
\hline
%Q0151+048A    & 1.922& 1.934 &  0.93&      & 20.36	  &           &  -	 &		   & [1]     \\ [removed proxim.]
%2233.9+1318   & 3.298& 3.150 &  $2.5\pm0.1$   & 19.5 & 20.00	      &	5.9      &UV	     & $-0.80\pm 0.24$ & [8,9]   \\
Q0139-0824$^2$ & 3.017& 2.677 &  $1.60\pm0.05$  & $13.0\pm0.4$& $20.70\pm0.15$&  -	&    -      & $-1.15\pm 0.15$ & [16]	\\
Q0338-0005     & 3.068& 2.230 &  $0.49\pm0.12$ & $ 4.1\pm1.0$& $21.05\pm0.05$&  -	&    -      & $-1.25\pm 0.10$ & [16]	\\
PKS0458-02     & 2.286& 2.039 &  $0.31\pm0.04$ & $ 2.6\pm0.3$& $21.77\pm0.07$& $>$1.5	&Ly$\alpha$ & $-1.19\pm 0.10$ & [3,6,14]\\
PKS0528-250    & 2.797& 2.811 &  $1.14\pm0.05$ & $ 9.1\pm0.4$& $21.27\pm0.08$& 4.2	&Ly$\alpha$ & $-0.75\pm 0.10$ & [4,5,9,16] \\
Q0918+1636$^1$ & 3.07 & 2.583 &  $1.98\pm0.01$ & $16.2\pm0.1$& $20.96\pm0.05$& $8\pm3$  &H$\alpha$  & $-0.12\pm 0.05$ & [11,12] \\ 
Q0953+47$^2$   & 4.457& 3.404 &  $0.34\pm 0.10$& $ 2.6\pm0.8$& $21.15\pm0.15$&  -	&     -     & $-1.80\pm 0.30$ & [16]	\\
J1135-0010     & 2.89 & 2.207 &  $0.10\pm 0.01$ & $ 0.8\pm0.1$& $22.10\pm0.05$&  $25\pm6$&H$\alpha$  & $-1.10\pm 0.08$ & [13]	\\
Q2206-1958a    & 2.559& 1.920 &  $0.99\pm 0.05$& $ 8.5\pm0.4$& $20.65\pm0.07$& 5.7	&UV	    & $-0.54\pm 0.05$ & [7,9]	\\
Q2206-1958b    & 2.559& 1.920 &  $1.2\pm0.1$   & $10.3\pm0.9$& $20.65\pm0.07$& 4.2	&UV	    & $-0.54\pm 0.05$ & [2,9]	\\
Q2222-0946     & 2.926& 2.354 &  $0.8\pm0.1$   & $ 6.7\pm0.8$& $20.65\pm0.05$& $9.5\pm1$&H$\alpha$  & $-0.46\pm 0.07$ & [8,10,14,15] \\
HE2243-60      & 3.01 & 2.328 &  $2.80\pm0.20$ & $23.4\pm1.7$& $20.62\pm0.05$& $18\pm2$ &H$\alpha$  & $-0.72\pm 0.05$ & [1,16,17]   \\
\hline  									
\end{tabular}
\flushleft References: %[1] \citet{fyn99};
[1] \citet{bou13}; [2] \citet{wea05}; [3] \citet{mol04}; [4] \citet{mol93}; 
[5] \citet{mol98}; [6] \citet{wol05}; [7] \citet{mol02}; [8] \citet{per12}; [9] \citet{chr14}; [10] \citet{fyn10}; [11] \citet{fyn11}; 
[12] \citet{fyn13}; [13] \citet{not12b}; [14] \citet{kro13}; [15] \citet{jor14}; [16] \citet{kro12}; [17] \citet{bou12}.  \\
\flushleft Notes: $^1$In the same sightline, \citet{fyn13} reported the detection of [OIII] emission line associated to 
a second DLA at $z = 2.412$. $^2$This DLA has been reported by \citet{kro12}, 
citing work in preparation by other authors.  Errors on the impact paramaters are
from \citet{kro12}. We do not include in this list the proximite DLA Q0151+048A \citep{fyn99} and the 
sub-DLA 2233.9+1318 from \citet{djo96}.\\
\end{table*}

Specifically, the number of random galaxies inside an annulus $\Delta b$ at a given 
projected separation $b$ from a DLA is 
\begin{equation}
N_{\rm ran} = 2\pi b \Delta b \mu_{\rm ran}\:,
\end{equation}
where $\mu_{\rm ran} = (2.566\pm0.046)\times 10^{-4}~\rm kpc^{-2}$ 
is the surface number density of galaxies at the depth of our imaging survey, which 
we compute using the previously defined apertures at large distances from the quasar positions. 
By comparing $N_{\rm ran}$ to the observed number of galaxies $N_{\rm obs}$ in a given annulus around the 
DLAs, we obtain a binned estimator of the projected DLA-galaxy correlation function, $w_{\rm dg}(b)$, defined by
\begin{equation}
w_{\rm dg}(b,\Delta b) = N_{\rm obs}(b,\Delta b)/N_{\rm ran}(b,\Delta b) -1\:.
\end{equation}
Errors are simply set according to Poisson statistics in the observed number counts.
A similar analysis, but employing mock galaxy catalogues constructed in apertures of 200 kpc 
and the DD/RR estimator, yields the same result. 
Given our qualitative discussion, we do not resort to the more advanced techniques for measuring 
correlation functions and estimating errors, which exist in the literature \citep[e.g.][]{lan93}.

The measurements of $w_{\rm dg}(b)$ are shown in Figure \ref{fig:chidla}. In agreement with the nearest neighbor 
statistics presented in Section \ref{impactb}, the projected DLA-galaxy correlation function exhibits an excess for 
the innermost bin, in the interval $0-20~\rm kpc$. This excess is however
only marginal, especially since our simple errors probably underestimate the true variance. 
Furthermore, this excess is severely dependent on the choice of the bin size, as is 
typical for correlation functions \citep[e.g.][]{cro97}. Increasing the innermost bin size to 30 kpc would cause the signal 
to disappear due to the rapid increase in the number of background sources, 
as one could also guess from Figure \ref{fig:bsfr}.

Conversely, for $b_{\rm dla}> 20~\rm kpc$, 
we do not see any excess of galaxies in proximity to the DLAs compared to background values. 
This is expected, since we are computing the projected correlation function without spectroscopic information,
thus integrating over a huge volume.  We can gauge the sensitivity of our measurement to DLA host galaxies
by recomputing the correlation functions after injecting 32 or 16 fake DLA galaxies in the observed 
catalogues with impact parameters $b_{\rm dla} = 20-100~\rm kpc$.  In this simple experiment, 
we assume that all galaxies in the observed catalogues are random interlopers, and that the 
fake sources are the only galaxies associated to the DLAs. Results from this calculation 
are shown in Figure \ref{fig:chidla}. As expected, the presence of tens of 
DLA galaxies at these impact parameters boosts the clustering signal, although the detection 
remains marginal.  A comparison between $w_{\rm dg}(b)$ with mock and real data
suggests that a large fraction ($\gg 50\%$) of DLA galaxies with SFR $\dot\psi \ge 2$\sfr\ and $b_{\rm dla} = 20-100~\rm kpc$ are not present in our data, thus implying that 
not all DLAs originate nearby to highly star-forming galaxies.

\subsection{Comparisons with other DLA searches}

The search for DLA host galaxies has been a long-time effort which, 
especially thanks to recent new detections \citep[e.g.][]{fyn11,per11,per12,not12b,fyn13,bou13,jor14},
has yielded 11 DLA galaxies currently known at $z> 1.9$. In Table \ref{known},
we present a summary of the properties of these confirmed DLA galaxies.
This compilation is the result of several heterogeneous searches, which include 
both serendipitous discoveries, and results from programmes that have pre-selected 
targets according to absorption properties, such as high metallicity. 
For this reason, it is quite difficult to reconstruct the selection function of these 
searches, and it is even harder, if not impossible, to obtain a full census of the DLAs for 
which a search has been attempted without success. Some notable exceptions are
the recent VLT/SINFONI survey by \citet{per11} and \citet{per12}, or the Gemini/NIFS 
programme by Wang et al. (MNRAS submitted), as these authors report both detections and 
non-detections.

\begin{figure}
\includegraphics[angle=90,scale=0.32]{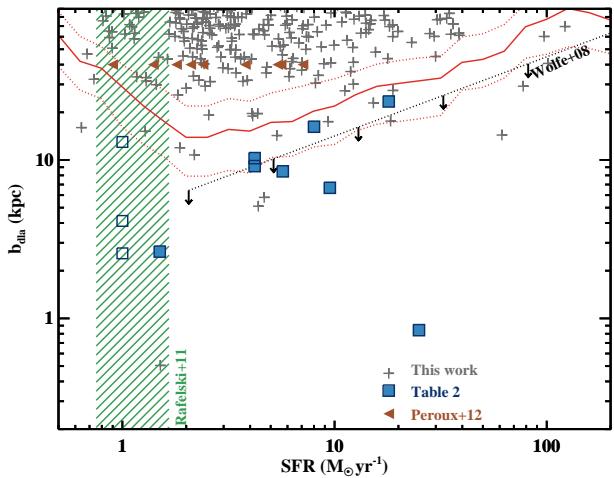}
\caption{Distribution of impact parameters as a function of SFRs for the 
detected galaxies in proximity of the DLA position (grey crosses), compared to the 
distribution of the confirmed DLA galaxies from Table \ref{known} (blue squares). 
Empty squares indicate galaxies without published SFRs, which are shown at the
arbitrary value of 1\sfr. The brown triangles mark the upper limits on the SFRs 
for the 9 DLAs without counterparts in the sample of \citet{per12}, which we plot at the maximum distance 
probed by their observations. The red solid and dotted lines represent the locus of the minimum 
impact parameters and the corresponding  $25-75$ percentiles (as in Figure \ref{fig:bsfr}). 
The black line and upper limits show the maximum impact parameter as a function of 
SFR for a compact LBG that would satisfy the SFR surface density 
inferred for the high-cool population. The green dashed region shows instead the $25-75$ percentiles 
of the SFR distribution for the compact LBGs included in the stack by \citet{raf11}.}\label{fig:otherdlas}
\end{figure}

\begin{figure}
\includegraphics[scale=0.43]{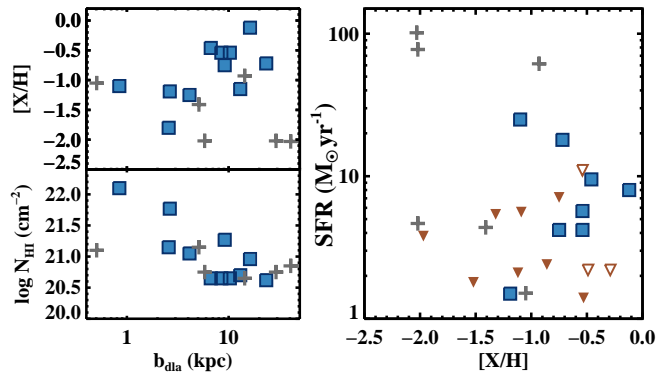}
\caption{{\it Left panels.} 
The metallicity (top) and column density (bottom) distributions of known DLA galaxies 
(blue squares) compared to the distribution of the sources detected in 
our survey (grey crosses) for $\dot \psi \ge 1~$\sfr\ and $b_{\rm dla} \le b_{\rm min, 25\%}$. 
{\it Right panel.} Same as the central panels, but comparing the SFRs in emission with the 
metallicity in absorption. The limits from \citet{per12} and Wang et al. (submitted) 
are also included as downward triangles (filled and open, respectively).  
SFRs are shown as published, and thus these measurements contain 
additional scatter due to inhomogeneous 
tracers and calibrations.}\label{fig:otherdlasbis}
\end{figure}

A comparison between the properties of the galaxies listed in Table \ref{known}
and the results of more homogeneous searches, such as our study, becomes therefore critical to 
understand the extent to which the confirmed DLA galaxies are representative of the 
general population of DLA hosts. This comparison is offered in Figure \ref{fig:otherdlas}, where we examine 
the distributions of impact parameters, metallicity, column densities, and SFRs for 
the known DLA hosts, the galaxies detected in our images, and the results of the IFU searches. 
Based on the published star formation indicators for $8/11$ DLA galaxies (Table \ref{known}),
it appears that the population traced by the known objects is generally above 
the completeness limit of our survey ($\dot\psi = 2~$\sfr). In the following, we will
assume that the three DLAs without published SFRs (Q0139$-$0824, Q0338$-$0005, and Q0953$+$47) lie 
within the sensitivity of our images. We note, however, that the compiled SFRs in Table \ref{known}
have a substantial uncertainty due to observational errors (e.g. slit losses in some cases).
They are also quite heterogeneous, due to the use of different SFR indicators and 
calibration schemes in the literature.

With these caveats in mind, Figure \ref{fig:otherdlas} shows that the known DLA galaxies 
lie all at impact parameters below the typical minimum impact parameter for a random background 
population, with $\le 8/11$ systems having impact parameters below
$b_{\rm min, 25\%}$\footnote{The inequality arises from the fact that the DLA galaxy 
Q0139$-$0824 at $b_{\rm dla}=13.0 \pm 0.4$ kpc may intersect 
the region $b_{\rm dla} > b_{\rm min, 25\%}$ for $\dot\psi \sim 1.5-7$\sfr.}. 
Conversely, our survey indicates that $\le 6/32$ of the DLA galaxies are found below 
$b_{\rm min, 25\%}$ at comparable SFRs. In our survey 
no more than $\sim 15\%$ of the DLAs have hosts with $\dot\psi \gtrsim 1$\sfr\ at low 
impact parameters. Conversely, previous searches would suggest instead that these SFRs and 
impact parameters should arise in $\sim 70\%$ of the DLAs. This comparison is affected by the small 
sample size, but it highlights how the emission properties of the confirmed hosts cannot be generalized 
to the entire DLA population without considering the various selection effects 
for the sample listed in Table \ref{known}.

Similar considerations can be made from the
VLT/SINFONI survey by \citet{per12}. This programme provides IFU data and hence 
a complete spectroscopic search up to distances of $\sim 40~\rm kpc$
in 10 $z \ge 2$ DLAs with a wide range of absorption properties,
similar to those shared by our targets.  A direct comparison between the SINFONI survey and known 
DLA galaxies is slightly complicated by the wide range of sensitivities 
reached by this programme, with upper limits between $\dot\psi \sim 1-10$\sfr.
Nevertheless, the low detection rate (a single detection in a sample of 10 DLAs) reinforces
our conclusions \citep[see also][]{bou12}. Similarly, Wang et al. (submitted)
did not find DLA hosts in their Gemini/NIFS IFU data down to $\dot\psi \le 2.2$\sfr\ for two systems, 
and $\dot\psi \le 11$\sfr\ for a third.

As discussed, it is far from trivial to quantify the relative importance of
the different selection effects, and here we bring to the attention of the 
reader three of the most obvious ones. First, the sample listed in Table \ref{known}
suffers from a luminosity bias, which is typical of any imaging or spectroscopic 
search. In this case, the confirmed DLA galaxies would represent only the ``tip of the iceberg'' 
at the bright end of the luminosity function. The luminosity bias alone could explain the 
excess of sources at small impact parameters reported in Figure \ref{fig:otherdlas}. 
However, it is worth considering a second more subtle 
selection effect which may arise from the observational technique adopted in the discovery of 
several of the known DLA hosts. A non-negligible number of these DLA galaxies have been 
discovered by triangulating the signal detected in two or three slits centered at the quasar position 
\citep[see e.g. figure 1 in][]{fyn10,mol04}. In this layout, the likelihood of detecting 
galaxies at small impact parameters is higher, purely because this is where the
area covered by the slits is maximal. While the majority of the DLAs would be detected 
given the model proposed by \citet{fyn08}, a fraction of DLA hosts may still be undetected purely 
because they lie at larger impact parameters. This effect 
would skew the impact parameter distributions of the detected DLAs towards small $b_{\rm dla}$, 
similarly to the observations in Figure \ref{fig:otherdlas}. 

Furthermore, as highlighted in Figure 
\ref{fig:otherdlasbis}, the absorption properties of the DLAs with known hosts are skewed towards 
high metallicity (right panel), and slightly towards higher column densities (left bottom panel). 
The excess of systems at high metallicity is a known consequence of the well-documented 
targeting strategy. Observations of metal-rich DLAs are indeed favored because of a possible 
mass-metallicity relation \citep{mol04,fyn08,fyn10,nee13} that would increase the odds of detecting 
the hosts of the more metal-rich systems.  
From this discussion, we conclude that: i) care should be exerted when interpreting the properties 
of DLA host galaxies solely based on compilations like the one in Table \ref{known}; ii) models 
aiming to reproduce DLA galaxies should not only reproduce the scaling relations of the 
known DLA hosts, but also the number of non-detections in homogeneous searches 
(such as the present one).

Finally, in Figure \ref{fig:otherdlasbis}, we also compare the few galaxies 
detected in our survey at $b_{\rm dla} \le b_{\rm min, 25\%}$ against some previously reported 
correlations between absorption and emission properties in DLA hosts. 
Correlations between impact parameters and column densities or metallicity have been already 
reported for the known DLAs \citep[e.g.][]{kro12}, as is also shown in the panels of 
Figure \ref{fig:otherdlasbis}. Our candidates appear instead to be more scattered, 
but these deviations are not particularly significant in the absence of
spectroscopic confirmation. When examining how SFRs vary with metallicity
(right panel of Figure \ref{fig:otherdlasbis}), no clear trend is evident in 
either our sample of candidate hosts or in the sample of confirmed DLA galaxies. 
To first order, if a mass-metallicity relation is in place 
for DLAs \citep{mol04,fyn08,nee13,mol13}, one would expect a similar correlation  
between SFRs and metallicity, given the known correlation between SFRs and masses
in galaxies. However, due to the small sample size and the intrinsic 
errors in the SFRs mentioned above, this lack of correlation is  inconclusive.
This is especially true for our candidates without redshift confirmations.  
Furthermore, the trend of metallicity with impact parameter complicates the relation
between metallicity in absorption and SFR integrated over the entire host galaxy  \citep{fyn08}.
It would nevertheless be interesting to augment the size of uniformly-selected DLA hosts to 
investigate further how a mass-metallicity relation for DLAs is reflected 
into a metallicity-SFR correlation. 
 
\subsection{Implications for cooling rates and emission from LBG outskirts}\label{dlagalcool}

Having characterized the distribution of galaxies in the surroundings of DLAs, 
we are now able to comment on the last possible scenario for the heating mechanism 
of the high-cool DLAs, which we discussed in Section \ref{sec:cool}. By means of the 
limits on the {\it in-situ} SFRs, we highlighted a potential discrepancy between 
the SFRs that would provide the heating rate necessary 
to power high-cool DLAs and the limits inferred by our direct observations 
for both extended low-surface brightness sources and compact ($r\lesssim 0.25''$) LBGs embedded in 
the DLA gas at very small impact parameters. As noted, this tension could be resolved if 
the heat is provided by LBGs at larger impact parameters, an hypothesis we can now 
address explicitly. 

Due to the $r^{-2}$ dependence of the 
UV flux, LBGs at larger distances have to be much more luminous to provide the 
same photon flux at the DLA position. In  
Figure \ref{fig:otherdlas}, we compare the locus of impact 
parameters as a function of SFRs for all the galaxies detected near
the targeted DLAs with the maximum impact parameters at which we can find an LBG
that satisfies the constraints imposed by the DLA cooling rates 
(cf. Figure \ref{fig:cool}). This limit is computed as follows. 
Given the size of LBGs $r_{\rm lbg}$, the size of the associated 
DLA $r_{\rm dla}$ (and hence of the maximum impact parameter) is constrained by the area 
covered by DLAs relative to the area covered by LBGs, $C_{\rm dla,lbg}$.
\citet{wol06} and \citet{wol08} estimate $C_{\rm dla,lbg}\sim 330$ between $z\sim 2.5-3.5$. We can 
therefore compute the SFR required to satisfy the SFR surface density 
inferred for the high-cool population as 
$\dot\psi=\Sigma_{\rm sfr} \pi r_{\rm dla}^2 = \Sigma_{\rm sfr} \pi r_{\rm lbg}^2 C_{\rm hc}$, 
where the covering fraction of the high-cool population is $C_{\rm hc} \sim 1/2 C_{\rm dla,lbg}$,
as discussed in \citet{wol08}.

Figure \ref{fig:otherdlas} shows the resulting limits on the impact parameters 
for LBGs, where the minimum at $\dot\psi \sim 2~$\sfr\ is imposed by the constraint  
$r_{\rm lbg} \ge 0.5~\rm kpc$. In this calculation we further assume 
$\Sigma_{\rm sfr} \sim 10^{-1.8}~$\Ssfr\ for the high-cool DLAs from the median of the distribution 
shown in Figure \ref{fig:cool}. We note that the presence of a dust-obscured population, 
unexplored in our study, would not contribute towards the budget of UV photons needed to satisfy 
the model prediction for the high-cool DLAs, as these photons will be screened by the same dust 
that prevents the detection of the galaxies in our imaging.
 
Given our sample of $32$ DLAs and assuming a roughly equal split between low-cool and high-cool DLAs, 
we would expect to have detected approximately 8 galaxies below the impact parameter limits 
imposed by the calculation above. Our data show that $\le6$ candidate DLA hosts 
(without double-counting the two  detections in 5:G5 field) are found 
below this line. If we allow for possible leakage in the 10:G11 field, 
only $\le 5$ galaxies are found out of the $\sim 8$ galaxies expected. 
Currently, these numbers are in formal agreement, but the modest number of 
star-forming galaxies below $\sim 30~\rm kpc$ may become uncomfortable for the model of 
\citet{wol08} if some of these candidates were found to not be at the 
DLA redshift. However, any tension should be considered modest, if present at all, given the 
many uncertainties at play. Firstly, this comparison hinges on a small number of 
galaxies; secondly, the fraction of high-cool DLAs relative to low-cool 
DLAs is unknown in our sample and may differ from what has been reported by 
\citet{wol08}; thirdly, previous estimates of the relative covering fraction of DLAs and
LBGs rely on shallower luminosity functions than those now available 
\citep[e.g.][]{ala14}. It will be very interesting to use larger samples and new direct 
measurements to investigate the assumptions at the base of the model by \citet{wol03,wol03a}.

Finally, with a census of the LBGs in the surroundings of these DLAs, 
we can compare our findings with the model proposed by \citet{raf11}, according 
to which DLAs arise in the outskirts of compact LBGs. In Figure \ref{fig:otherdlas}, we show the 
$0.25-0.75$ percentiles of the SFRs for the galaxies included in the stack of \citet{raf11},
which lie in the range $\dot\psi = 0.8-2~$\sfr. Unfortunately, our survey starts being incomplete
below $\dot\psi = 2~$\sfr, preventing a statistical comparison between our detections and 
the presence of compact LBGs. Again, deeper imaging surveys will provide valuable 
constraints for this model. 

\section{What are DLA\lowercase{s}?}\label{other}

A definite answer to the question of what DLAs are has been hard to obtain, despite 
much progress in characterizing the emission and absorption properties of these absorbers.
Besides the interest in unveiling the nature of these systems, a
deeper understanding of DLAs has far reaching implications. DLAs represent 
the major repository of neutral gas in the $z\sim 2-3$ Universe, 
and studies of the SFRs in representative samples of these absorbers, like the one 
presented here, offer new insight into how the most significant HI overdensities 
relate to star formation.

The primary results emerging from our analysis are: 1) new 
limits on the {\it in-situ} SFR of DLAs, to $\dot\psi < 0.09~$\sfr\ for compact regions or
$\dot\psi < 0.27~$\sfr\ for more extended sources; 2) a deep search for galaxy counterparts 
in the surroundings of DLAs, through which we have detected 
an excess of sources only at small impact parameters ($\lesssim 30~\rm kpc$), and 
for $\le 4/32$ DLAs to a completeness limit of $\dot\psi = 2~$\sfr.
Thus, our survey places new quantitative limits on the allowed parameter space
for DLA galaxies, ruling out the parameter space 
$\dot\psi \ge 2~$\sfr\ and $b_{\rm dla} < b_{\rm min,25\%}\rm ~kpc$ 
for the general DLA galaxy population. 

In this section, we aim to discuss how these new limits fit with the 
empirical knowledge accumulated over the years on the origin of DLAs.
In particular, we will focus on three scenarios: i) DLAs as 
systems where star formation occurs only throughout the absorbing gas; 
ii) DLAs as moderately star-forming galaxies which are however sufficiently dusty to
become invisible at UV wavelengths; iii) DLAs as UV-bright star-forming galaxies, or LBGs. 
At this stage, the discussion will be qualitative and, occasionally, even speculative.
Future theoretical work is encouraged to incorporate more quantitatively 
the new empirical constraints we have provided in the context of the many models of DLAs that 
can be found in the literature 
\citep[e.g.][]{fyn99,nag07,fyn08,pon08,tes09,hon10,cen12,rah14,bir14,bar14,dan14}.

Furthermore, in the following, we will not explore the even harder question of the morphology
of DLAs, which can take the form of disk-like structures, clumps, tidal debris, inflows, outflows, 
or even quite extended structures \citep{gio89}.
We will simply focus on the more general relationship between DLA gas and star formation. 
It is also implicit that a rigorous classification within the three classes mentioned above 
is not mandatory, or may even be incorrect if DLAs arise from a diverse population.

\subsection{DLAs as aggregates of gas and stars}\label{sec:lsb}

In this first class, we envision DLAs as aggregates of \HI\ gas giving rise to 
the absorption signature, with star formation occurring throughout the same gas. 
In this scenario, the DLA gas and the DLA galaxy are coincident in the same system,
which can be probed either in absorption (the DLA gas) or in emission (the DLA galaxy). 
The plausibility of this scenario has already been investigated in the literature,
most notably by \citet{wol06}, or more recently by means of composite spectra 
of DLAs \citep{rah10,not14,cai14}. In agreement with these previous studies, our direct measurements 
of the {\it in-situ} SFRs via FUV emission rule out that typical DLAs are star-forming 
pockets of gas that are buried under the glare of the background quasars, up to SFR surface 
density limits of $\Sigma_{\rm sfr} = 10^{-1.54}~$\Ssfr\ and $\Sigma_{\rm sfr} = 10^{-2.62}~$\Ssfr, 
for sizes of $2~\rm kpc$ and $12~\rm kpc$, respectively (Figure \ref{fig:sflaw}).

During our analysis, we emphasized that the low {\it in-situ} SFRs inferred from our observations 
are not inconsistent with other observed properties of DLAs, essentially allowing for 
a DLA population with typically very little star formation.
As shown in Figure \ref{fig:metal}, even modest star formation 
events can still satisfy the local metal enrichment.
Additionally, provided that most of the DLA gas at higher 
redshift merges onto other structures that are forming stars at higher rates \citep{pro09}, 
we can still preserve the general argument that DLAs contain enough neutral hydrogen to form about 
a third of the stars produced in the universe by $z\sim 0$, even without imposing that these 
stars are formed {\it in situ}. 

Therefore, our observations do not exclude the possibility that star formation occurs
throughout the DLA gas at rates below our inferred limits, leaving open 
a scenario in which stars are formed fully within the DLA gas for the origin of most DLAs.
However, as argued in Section \ref{ins:sflaw}, this scenario is more difficult to 
corroborate from a theoretical point of view: typical DLAs with column densities 
$N_{\rm HI} = 10^{20.3}-10^{21}~\rm cm^{-2}$ may not provide a favorable environment for star formation.
Direct observations are needed to test this hypothesis, but we note that, in line with our 
argument, there is evidence for {\it in-situ} star formation only in DLAs with 
$N_{\rm HI} > 10^{21}~\rm cm^{-2}$ \citep{raf11,not14}, and not at lower column densities. 
Furthermore, the potential discrepancy between the observed limits and the inferred SFR surface 
densities from the \CIIs\ cooling rates (Figure \ref{fig:cool}) may be resolved by associating 
``high-cool'' DLAs to nearby star-forming galaxies \citep{wol08}, a fact that would make 
a picture in which DLAs form stars fully within the absorbing gas insufficient to account 
for all the known observables. 

And while recent studies have uncovered a population of ``dark galaxies'' \citep{rau08,can12},
despite the evidence that DLAs lack appreciable star formation {\it in situ},
we believe that it would be incorrect to extrapolate our findings to conclude that 
DLAs simply arise from systems not related to star formation.
In fact, the general DLA population 
is not an aggregate of pristine gas, nor does it share the same metal content of the 
IGM \cite[e.g.][]{raf12}. Furthermore, the median velocity widths of DLAs \citep{pro97} 
suggest that DLA gas resides in significant potential wells, in which it is plausible to expect 
star-forming regions. 

We conclude that a scenario in which DLAs are aggregates of gas and 
stars is not completely ruled out by observations, although it is not the most favorable hypothesis, 
given the aforementioned issues. Thus, if not fully embedded with DLA gas, star formation has 
to occur at least in the proximity of DLAs, a scenario we will explore in the next two sections. 

\subsection{DLAs as dusty star-forming galaxies}\label{dustobs}

In the two scenarios we discuss in this and in the following section, we consider 
DLAs that are associated with some star formation, which is however not fully 
embedded in the absorbing gas, as it was instead the case for the scenario in Section \ref{sec:lsb}. 
The readers who are familiar with the work by \citet{wol06} and \citet{wol08} will recognize this 
as the ``bulge hypothesis''. 

A substantial fraction of the total cosmic star formation arises from obscured regions 
\citep[e.g.][]{bla99,cha01,bow09,dec14}, and given that DLAs contain enough neutral hydrogen to 
account for about a third of the stars ever formed, we consider whether most DLAs may originate 
in proximity to dusty star-forming regions. 
In this category, we do not envision associations between DLAs and
submillimetre or ultra luminous infrared galaxies for a few reasons. 
Firstly, the space density of these extreme populations is insufficient to account for 
the more abundant DLA population. Secondly, the few direct searches in emission 
for molecular gas in DLAs rule out the presence of massive reservoirs of cold gas and dust 
near the absorbing gas \citep[e.g.][, Dessauges-Zavadsky et al. in prep.]{wik94}.

Thus, with the expression ``dusty star-forming galaxies'', we identify instead a population 
of galaxies with masses comparable to or even lower than the classic LBGs 
($M_{\rm halo} \sim 10^{11}-10^{12}~\rm M_\odot$), in which star formation occurs within regions 
that are dusty enough to be attenuated and become invisible to surveys that select 
galaxies solely based on FUV emission.  Several studies \citep[e.g.][]{red04,red10} concluded that 
dust in bright LBGs suppresses the FUV emission by a factor of $\sim 4-5$. 
Extinction up to $\sim 2$ mag, if extrapolated also to fainter systems, 
could therefore play an important role in shaping the statistics of the impact parameter as a 
function of SFR (Figure \ref{fig:bsfr}), especially given that our results
are based on rest-frame wavelengths $\lesssim 1000-1100$\AA.

Therefore, the presence of a relatively massive but dusty galaxy population near DLAs 
should be considered further, as it would for instance ease the potential tension between studies 
that reveal how DLAs originate from regions with high bias \citep{coo06,fon12} and theoretical 
works that instead struggle to incorporate both this high bias and all the other observed DLA 
properties in a single model \citep[e.g.][]{bir14,bar14}.

However, measurements of the {\it in-situ} metallicity 
(e.g. Figure \ref{fig:metal}) place DLA gas in regions of modest dust content. This is
also seen directly in the low reddening measurements along the line of sight to quasars with DLAs 
\citep{vla08,kha12}, although we cannot fully rule out with current observations the presence of 
dusty clumps in close proximity to, but not overlapping with, the locations probed in absorption. 
Similarly, the paucity of cold gas and molecules in high-redshift DLAs 
\citep[see e.g.][]{car96,kan06,cur06,jor06,not08,ell12,sri12,kan14,jor14b} points towards an environment 
that is atypical for dusty, gas rich, galaxies. Finally, we note that, although the exact relationship 
between the observed cooling rate and the UV photon flux may be subject of future revision, the presence 
of high-cool DLAs disfavor a scenarios in which the UV radiation field is completely suppressed by dust.

In summary, given these contrasting pieces of evidence, the possibility that dust obscuration 
plays a role in shaping the detection rates of DLA galaxies remains open. 
This scenario is currently almost completely unexplored for galaxies of modest IR luminosity, 
and unbiased surveys at infrared and millimeter wavelengths to very faint levels are now 
needed to address the importance of dusty galaxies of modest mass as an important or 
even dominant population of DLA hosts.

\subsection{DLAs as UV-selected star-forming galaxies}

As already noted, the absorption properties
of DLAs (e.g. metallicity and cooling rates), the fact that DLAs 
contain most of the neutral gas reservoir at high redshift, and the 
results of cross-correlations of DLAs and LBGs suggest that DLAs 
arise in the vicinity of star forming galaxies. 
With the caveat of a possible dusty population discussed in Section \ref{dustobs}, 
it is therefore natural to associate DLAs with LBGs, as done by many authors in the 
literature \citep[e.g.][]{fyn99,wol03,mol04,fyn08,pon08,raf11,rah14}.
Given that DLAs are selected purely based on gas cross-section and not 
luminosity, and because of the difficulties of identifying DLA host galaxies, 
it seems natural to associate DLAs with the faint end of the luminosity function 
\citep{fyn08,rau08,raf11}. In this scenario, which is also corroborated by simulations 
\citep{hae00,nag07,pon08,rah14}, the few known DLA hosts with $\dot\psi \gtrsim 1-2~$\sfr\ 
arise from the brighter LBGs.

Our results are consistent with this picture, providing for the first time 
quantitative limits on the bright end of the luminosity function of the DLA host galaxies. 
Our observations rule out the possibility that the DLA galaxies are LBGs with 
$\dot\psi \ge 2~$\sfr\ (unobscured) which are outshined by the background quasars. 
Furthermore, the lack of LBGs with $\dot\psi \ge 2~$\sfr\ and $b_{\rm dla} \le  b_{\rm min,25\%}~\rm kpc$ 
(Figure \ref{fig:bsfr}) rules out ``bright'' LBGs with large \HI\ disks 
or more amorphous structures as those predicted by modern simulations 
\citep[e.g.][]{dan14} of sizes $\sim 5-20~\rm kpc$ as the dominant population that 
gives rise to DLAs. At face value, this empirical fact disfavors models in which most of 
DLAs originate from extended rotating disks \citep{pro97,mal01}.

Given these considerations, we are therefore forced to conclude that DLAs arise in the surroundings 
of fainter galaxies, with typical SFRs of $\dot\psi \lesssim 2~$\sfr\ \citep[e.g.][]{rau08,raf11}.
If these dwarf LBGs are centrals of their dark matter halos (i.e. if they are field galaxies),
then it would entirely explain the lack of bright LBGs in the surroundings of DLAs 
(e.g. Figure \ref{fig:chidla}). 
However, a scenario in which DLAs are typically isolated dwarf LBGs 
may be difficult to reconcile with the measured DLA bias \citep{coo06,fon12}, which suggests 
instead a comparable clustering amplitude of DLAs and bright LBGs. 
As speculated by \citet{fon12}, this tension may be alleviated if DLAs arise from 
dwarf galaxies which are instead satellites of more massive (brighter) LBGs,
but a more quantitative analysis is now needed. 
Incidentally, we note that the idea of an extended DLA gas structure, 
in which sub-halos are embedded and which is larger than the typical \HI\ disk, 
is what motivated the simple picture of extended disks put forward by early models 
\citep{pro97,mal01} to explain the observed kinematics. And this picture 
gains also support from modern cosmological simulations, in which DLA gas extends inside or in 
proximity to massive dark matter halos, also encompassing smaller sub-halos \citep{rah14}. 

To muddle the picture,  however, simulations still struggle to reproduce correctly some of the most 
basic properties of DLAs, such as the redshift evolution of their number \citep[e.g.][]{bir14}. 
Also, the low rate of detections in the IFU observations of \citet{per12} and \citet{per11} out 
to 40 kpc from the DLA position may not be easy to reconcile with this idea, as one would expect 
to detect the bright LBGs at the center of the massive parent halos at these distances
in some instances. Perhaps more critical is the lack of significant excess in the DLA-LBG correlation 
function beyond 20 kpc, which is at odds with the idea of an abundant DLA population
in satellites of massive LBGs out to their virial radii (Figure \ref{fig:chidla}).

We therefore conclude that, while the idea of a link between LBGs and DLAs is 
sound from a qualitative point of view, more work is needed to quantitatively 
incorporate the properties of the observed DLA galaxies 
\citep[e.g.][]{kro12}, the limits of the bright end of the luminosity function
(Figure \ref{fig:bsfr}), the DLA bias, kinematics, and the presence of metals in high-ionization 
states \citep{leh14} in a coherent model. Deep and complete
redshift surveys in the surroundings of these DLAs are further needed to better constrain 
the importance of clustering of dwarf galaxies around massive LBGs in shaping the
properties of the DLA host galaxies. These searches should however account 
for the subtle biases induced by the stochastic nature of star formation at these low masses,
which shapes the detection rates of faint galaxies \citep{das14,dom14}.

\section{Summary and conclusions}\label{conclusion}

In this paper, we have presented results from an imaging survey of 32 quasar fields
hosting DLAs at $z\sim 1.9-3.8$, with a median redshift $z\sim 2.7$. 
By leveraging the double-DLA technique that allows us to image 
the FUV continuum of DLA host galaxies at all impact parameters, including the innermost regions
traditionally outshined by the background quasars, we have directly studied the 
{\it in-situ} emission properties of the DLA gas and the distribution of the
surrounding galaxies. We have also systematically compared our findings to the properties of DLAs inferred 
from absorption spectroscopy. Our main results are summarized below.

\begin{itemize}
\item[--] By positioning apertures of 2 kpc and 12 kpc centered on the DLA in each field,
  we have constrained the median {\it in-situ} SFRs of the DLA gas to be 
  $\dot\psi \lesssim 0.38~$\sfr\ and $\dot\psi \lesssim 0.65~$\sfr\ in individual systems. 
  With a stacking analysis, we have derived even tighter constraints on the median SFRs of DLAs, 
  $\dot\psi<0.09~$\sfr\ and $\dot\psi<0.27~$\sfr\ in the 2 kpc and 12 kpc apertures. 
  Converted into SFR surface densities, these limits become $\log \Sigma_{\rm sfr}<-1.54~$\Ssfr\ and 
  $\log \Sigma_{\rm sfr}<-2.62~$\Ssfr, ruling out the existence of appreciable star 
  formation embedded in the DLA gas. These limits reinforce the statistical conclusion of 
  \citet{wol06} and are in line with expectations borne from the current theoretical understanding 
  of the star formation law. 
\item[--] By comparing the limits on the {\it in-situ} star formation to the observed 
  metal content of DLAs in absorption, we have found that DLA gas can be enriched to the 
  observed levels in $\gtrsim 0.05-10$ Myr for compact DLAs and in $\gtrsim 1-300$ Myr for 
  extended DLAs. 
  Given the available time for the enrichment to occur ($\sim 2$ Gyr), 
  even lower rates of star formation ($30-500$ times lower than the measured upper limits) 
  would suffice to enrich DLAs to the observed values, without the need for external sources 
  of metals. It thus appears unlikely that star formation in or near DLAs underproduces the metals locked 
  in DLAs, in line with the fact that DLAs contain only a modest fraction ($\sim 1\%$) of
  all the metals produced in high redshift LBGs.
 \item[--] When compared to the inferred SFR surface densities required to 
   satisfy the \CIIs\ cooling rates of the ``high-cool'' DLA population, we find a tentative 
   discrepancy between our measured limits and the star formation required to heat 
   the DLA gas. Furthermore, the number of star-forming galaxies in the 
   surroundings of the DLAs with our deep imaging is slightly lower than the number expected 
   to satisfy the heating rates predicted by models. This tension is however only very
   marginal and should be investigated further in larger samples. 
 \item[--] By studying the distribution of impact parameters as a function of 
   the star formation rates for all the galaxies detected in proximity to the 
   DLA position, we have identified 6 possible fields with an excess of sources compared
   to expectations from a random galaxy background. Additional follow-up spectroscopy 
   is needed to confirm the redshift of these candidate DLA galaxies. 
\item[--]  At the completeness limit of our survey ($\dot\psi = 2$\sfr), we 
   find that  $\le 13\%$ of the DLAs lie at impact parameters 
   $b_{\rm dla} / {\rm kpc} \le (\dot\psi/{\rm M_\odot~yr^{-1}})^{0.8}+6$, in contrast with 
   samples of known DLA hosts which appear to suffer from selection effects in luminosity and/or
   impact parameters. Furthermore, we do not detect excess in the projected DLA-galaxy correlation 
   function between $20-100~\rm kpc$, ruling out an abundant population ($\gg 50\%$) of bright 
   DLA host galaxies at these separations.
\end{itemize}

Our study adds new direct constraints to the debate of what DLA galaxies are, albeit not 
solving this puzzle. Our new findings, combined with findings from other 
studies, suggest that DLAs cannot originate from aggregates of gas which are forming 
stars throughout to levels of $\sim 0.1-0.6$\sfr. Through indirect 
observations and theoretical arguments, we argued that a 
more plausible scenario, which has been already advocated for by many authors, puts DLA gas 
in proximity to star-forming galaxies, possibly UV-selected LBGs, although the presence 
of a bias induced by dust needs to be investigated further. 
While it appears evident that DLAs are not directly 
associated to bright LBGs and that dwarf galaxies are the more natural hosts, one could envision 
associations with either dwarf isolated galaxies, or dwarf galaxies which are clustered with more 
massive LBGs. The latter case may be required to reproduce the observed bias of DLAs, 
but it appears to be at odds with the paucity of bright LBGs found in previous IFU searches 
or as suggested by our measurement of the DLA-galaxy projected correlation function. 
Further modeling is now required to quantitatively explore how to incorporate our new constraints 
with earlier observations and models for the origin of DLAs.

This study further opens new prospects to finally unveil the nature of DLAs. 
Our survey has placed strong limits on 
the presence of star formation in galaxies close to the quasar sightline. An 
important open question that future redshift surveys should address  
is the distribution of  star-forming galaxies at large impact parameters to the sightline. 
Large integral field spectrographs, such as MUSE 
at VLT and KWCI at Keck, will soon provide this piece of information. 
Similarly, the deployment of the full ALMA array should enable the first glimpse 
into a potential dust obscured galaxy population invisible to UV studies, 
which may be associated with (a fraction of) the DLA population, but which would have been 
completely missed by our study. 

Furthermore, we highlighted on several occasions the potential of a deeper imaging survey 
which adopts the double-DLA technique at the basis of our study. 
A similar survey, but with an improvement by a factor of 10 in depth, would 
yield stringent constraints on the {\it in-situ} properties of DLAs, critical to investigate
the SF law for low metallicity gas and the mechanisms that are 
responsible for the enrichment and heating of the largest  
neutral hydrogen reservoir in the high-redshift Universe. Furthermore, 
these observations would provide 
unprecedented information on the clustering of faint dwarf galaxies around
DLAs, as as well as a direct discriminant between models for DLAs.
Thus, carefully designed surveys at current facilities and, in the future, 
observations at the 30 m telescopes
have the potential of finally answering some of the most challenging questions that have characterized
DLA studies in the past thirty years.

\section*{Acknowledgments}

We thank J. Hennawi, M. Neeleman, and M. Rauch for useful comments and suggestions on this work. 
We also thank J. Fynbo for providing a copy of the code to compute the models presented in 
\citet{fyn08}. We fondly remember discussions on the early stages of this 
work with Art Wolfe. MF acknowledges support by the Science and Technology Facilities Council 
[grant number  ST/L00075X/1]. JXP acknowledges 
support from NSF-AST 1109447 and 1109452 awards. MR acknowledges support from an appointment to the NASA Postdoctoral 
Program at Goddard Space Flight Center. NK acknowledges support from the Department of Science 
and Technology through a Ramanujan Fellowship. 
Support for Program number 11595 was provided by NASA through a grant (HST-GO-11595.001-A) 
from the Space Telescope Science Institute, which is operated by the Association of Universities for Research in 
Astronomy, Incorporated, under NASA contract NAS5-26555.
This work has been partially supported by NASA grant HST-GO-10878.05-A. 
For access to the data used in this paper, please contact the authors. 

%\appendix

\label{lastpage}

\end{document}